# EAST AFRICA IN THE MALTHUSIAN TRAP?
## A statistical analysis of financial, economic, and demographic indicators

*Andrey Korotayev & Julia Zinkina*

**Abstract.** *A statistical analysis of financial, economic, and demographic indicators performed by the authors demonstrates (1) that the main countries of East Africa (Uganda, Kenya, and Tanzania) have not escaped the Malthusian Trap yet; (2) that this countries are not likely to follow the "North African path" and to achieve this escape before they achieve serious successes in their fertility transition; (3) that East Africa is unlikely to achieve this escape if it does not follow the "Bangladeshi path" and does not achieve really substantial fertility declines in the foreseeable future, which would imply the introduction of compulsory universal secondary education, serious family planning programs of the Rwandan type, and the rise of legal age of marriage with parental consent. Such measures should of course be accompanied by the substantial increases in the agricultural labor productivity and the decline of the percentage of population employed in agriculture.*

The Malthusian trap[1] is a rather typical for pre-industrial societies situation when the growth of output (as it is accompanied by a faster demographic growth) does not lead in the long-range perspective to the increase in per capita output and the improvement of living conditions of the majority of population that remains close to the bare survival level (see, *e.g.*, Malthus, 1978 [1798]; Artzrouni & Komlos, 1985; Steinmann & Komlos, 1988; Komlos & Artzrouni, 1990; Steinmann, Prskawetz, & Feichtinger 1998; Wood, 1998; Kögel & Prskawetz, 2001; Korotayev, Malkov, & Khaltourina, 2006b; Korotayev & Khaltourina, 2006; Clark, 2007; Conley, McCord, & Sachs 2007; Korotayev *et al.*, 2011).

In complex pre-industrial societies the Malthusian trap was one of the main generators of state breakdowns and sociopolitical collapses frequently resulted in millions of deaths (see, *e.g.*, Korotayev & Khaltourina, 2006; Korotayev, Malkov, & Khaltourina, 2006*b*; Chu & Lee, 1994;

---
[1] Using the terminology of non-linear dynamics one can also denote it as *the low-level equilibrium attractor* (cp. Nelson, 1956).

Nefedov, 2004; Turchin, 2003, 2005*a*, 2005*b*; Turchin & Korotayev, 2006; Turchin & Nefedov, 2009; Usher, 1989; van Kessel-Hagesteijn, 2009; Korotayev *et al.*, 2011; for a recent case in East Africa see André & Platteau 1998)[2].

The global modernization, and first of all the increase in per capita productivity due to major technological advances in the recent centuries in conjunction with demographic transition secured the systematic excess of output growth rates over population growth rates, which allowed most social systems to escape the Malthusian trap (see, *e.g.*, Artzrouni & Komlos, 1985; Lucas, 1998; Galor & Weil, 1999; Kögel & Prskawetz 2001; Korotayev, Malkov, & Khaltourina, 2006*a*; Pereira, 2006).

However, these modernization processes started later in Sub-Saharan Africa than in the rest of the world; and even in the recent decades the Malthusian trap tended to produce state breakdowns in this region (see, e.g., André & Platteau 1998). What is more, some parts of Tropical Africa (and, first of all, East Africa) appear to be trapped in the Malthusian Trap till now (though contrary views continue to be expressed [see first of all Kenny, 2010]). This looks especially salient if we compare the long term per capita food consumption dynamics in North Africa, on the one hand, and the three main countries of East Africa (Kenya, Tanzania, and Uganda), on the other (see Fig. 1):

---

[2] Note that Malthus himself considered warfare (including, naturally, internal warfare) as one of the most important results of overpopulation (in addition to epidemics and famines). What is more, he regarded wars, epidemics, and famines as so-called "positive checks" that checked overpopulation in pre-industrial systems (Malthus 1978 [1798]). Thus, in pre-industrial societies bloody political upheavals frequently turned out to be a result of the respective social systems being caught in the Malthusian trap.

**Fig. 1.**   Average per capita food consumption dynamics (kcal per capita per day) in North and East Africa, 1961-2009

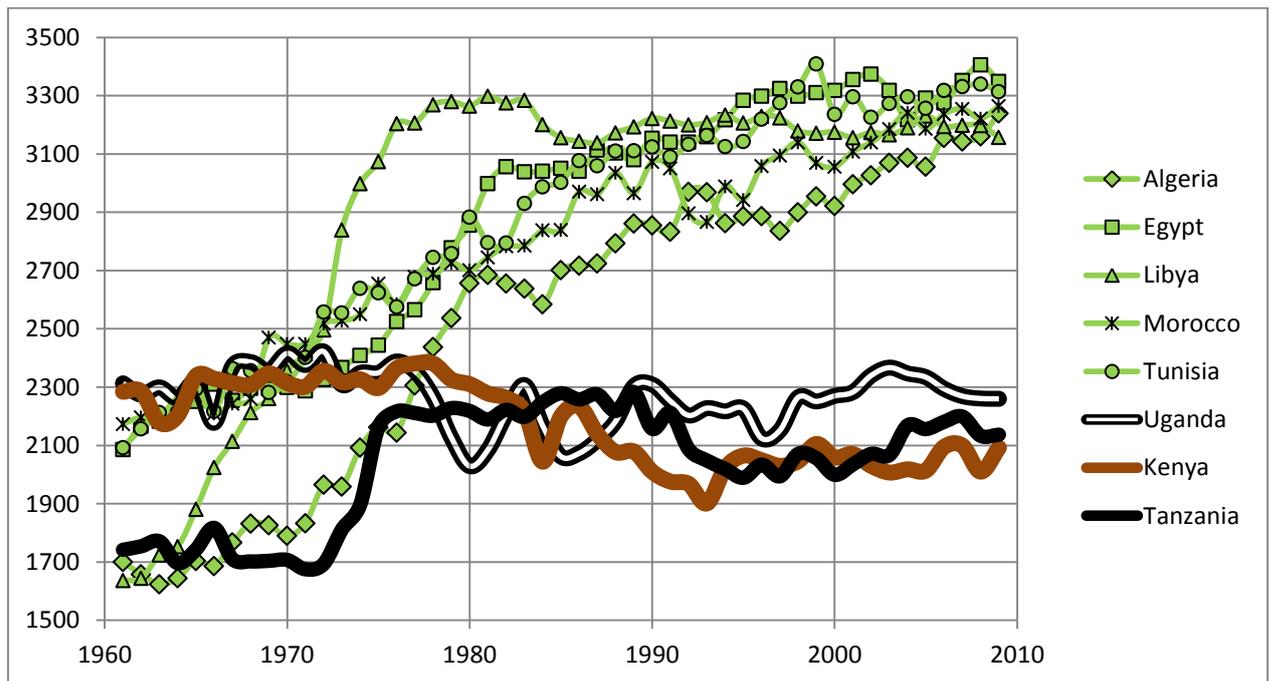

*Data source*: FAO 2013.

The WHO recommended norm of per capita is generally between 2300 and 2400 kcal/person/day (Naiken 2002), whereas 2100 is considered to be the minimum daily requirement (Joint FAO/WHO/UNU Expert Consultation 1985; FAO 2009: 24). Note also that according to Clark (2007) the average per capita calorie intake of about 2300 kcal/person/day or less is typical for those social systems that have not escaped the Malthusian Trap – such a level was observed, for example, in England and Belgium around 1800 (that is just before those countries started their escape from the Malthusian Trap – 2322 for Britain in 1787–1796 and 2248 for Belgium in 1812 [Clark 2007: 50]). Thus all the countries of North and East Africa appear to have still been in the early 1960s in the Malthusian Trap with the majority of their population remaining close to the bare survival level and had serious malnutrition problems. The situation was particularly bad in Libya, Tanzania, and Algeria, where the average per capita food consumption was even below the minimum daily requirement level; it was somehow better in Kenya, Uganda, Egypt, Morocco, and Tunisia[3]. However, it could hardly be called problem-free. The average per capita food consumption was still below the recommended and it was too close to the minimum daily requirement – which implied that a very substantial proportion of the population consumed below this level[4].

However, in the 1970-s – early 1980s all the countries of North Africa escaped rather successfully the Malthusian Trap and now these countries struggle more against overeating problems rather than with problems of undernourishment[5].

---

[3] Note that in the two East African countries of this group the situation was even better than in the North African countries.

[4] Note that all the countries that have firmly escaped the Malthusian Trap reached the level significantly higher than 2400 kcal/capita/day.

[5] For example, percentage of the obese among the modern Egyptians is one of the highest in the world (*e.g.*, Martorell *et al.* 2000; Korotayev, Zinkina 2011). According to Egyptian Demographic and Health Survey (conducted in 2008) 40% of Egyptian women and 18% of men were overweight because of overeating (Egypt Ministry of Health *et al.* 2009). According to a bit more recent data, this figures equal 22% for males and 48% for females (Badran, Laher 2011: 3).

This escape was accomplished first of all due to the fact that in this period all the countries of this region managed to achieve very high GDP growth rates that exceeded significantly the population growth rates, which resulted in very substantial increases in per capita GDP[6] (see Fig. 2):

**Fig. 2.** Relative Dynamics of Population, GDP, and GDP per capita in North African Countries during Their Escape from the Malthusian Trap

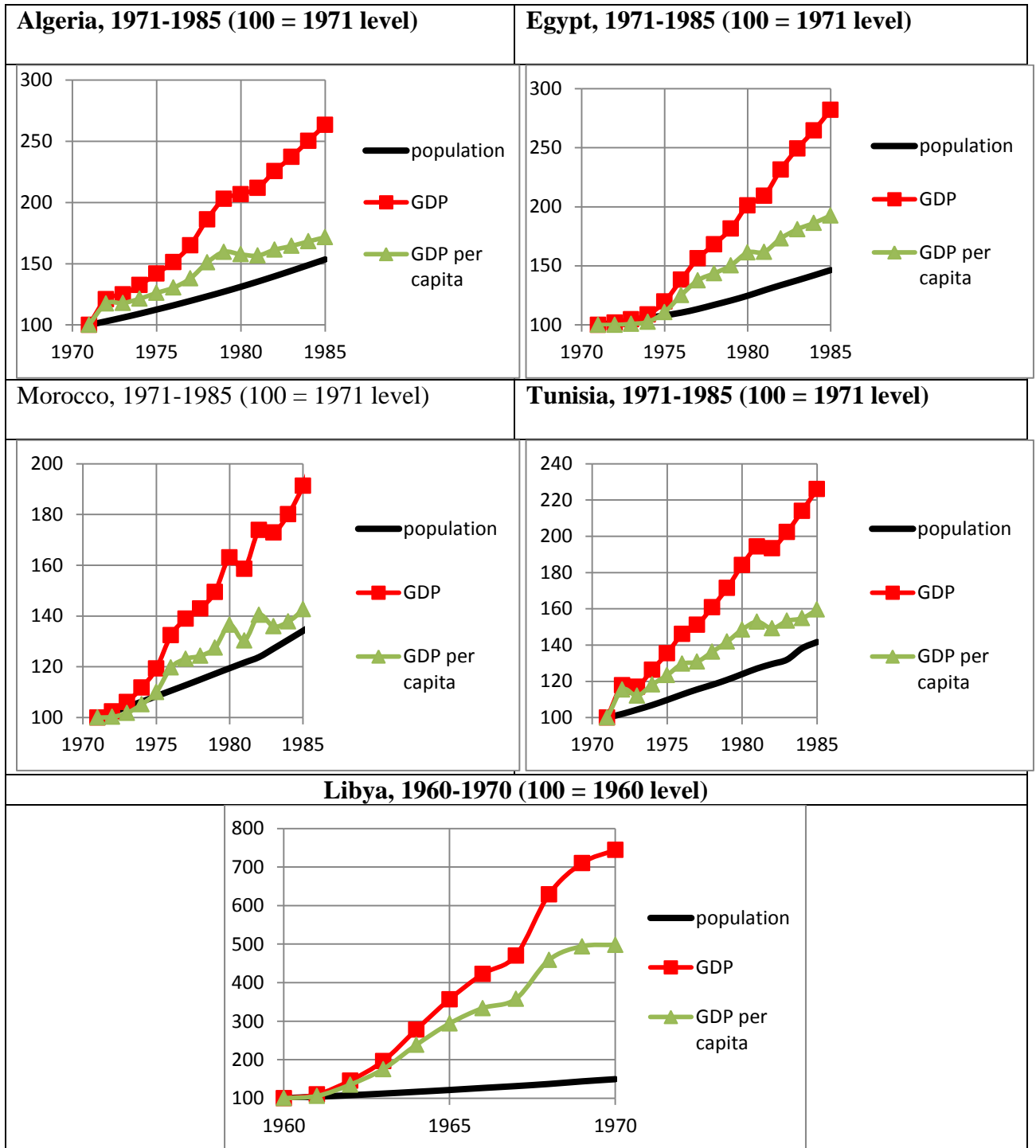

*Data source*: Maddison 2010.

This escape was further supported by the decrease of the fertility rates (see Fig. 3):

---

[6] Note that Libya made the respective breakthrough a bit earlier, in the 1960s.

**Fig. 3.** Dynamics of the Total Fertility Rate (TFR, births per woman) in North and East Africa (1960–2011)

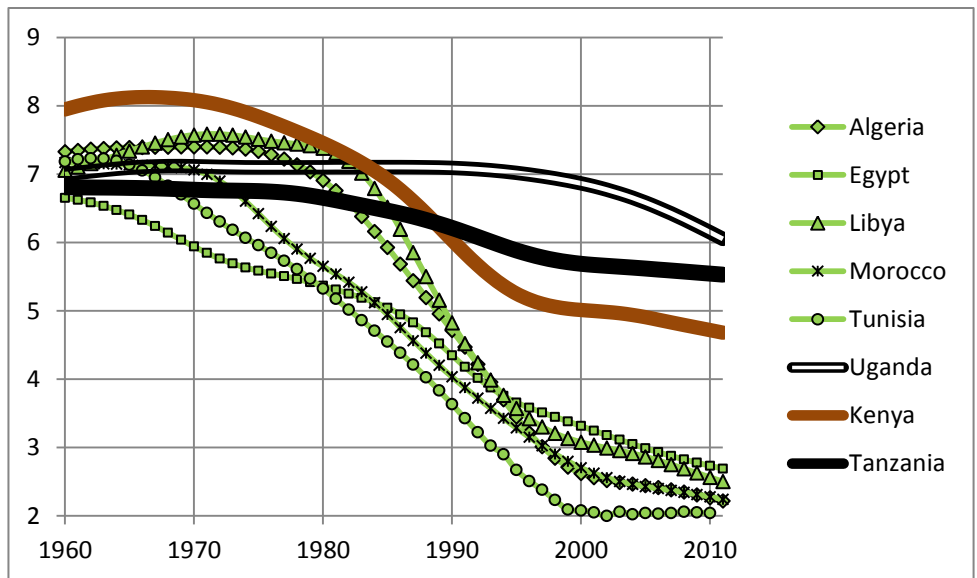

*Data source*: World Bank 2013: SP.DYN.TFRT.IN.

As we see, in this respect, North Africa also stands in sharp contrast with East Africa, where the total fertility rate remains very high till now.

Of course, in recent decades the TFR somehow declined in East Africa too, but the extent of this decrease was rather small, and it was mostly compensated by the decrease of the death rate (which was interrupted somehow by the HIV/AIDS epidemic, but resumed with its decline[7]). As a result, in a contrast with North Africa which observed a very significant decline of the population growth rates in the 1980s and 1990s, East African population growth rates remained mostly at such levels that were typical for North African countries before the start of the fertility decline in this part of the world (see Fig. 4):

---

[7] See, e.g., World Bank 2013: SP.DYN.LE00.IN, SP.DYN.IMRT.IN, SH.DYN.MORT, SP.DYN.CDRT.IN; Ngallaba *et al.* 1993: 71–78; Bureau of Statistics [Tanzania] and Macro International Inc. 1997: 97–104; Opiyo, Omolo, Imbwaga 2010; National Bureau of Statistics [Tanzania] 2000: 85–92; 2005: 123–130; 2011: 117–126; Kaijuka *et al.* 1989: 53–66; Statistics Department [Uganda] and Macro International Inc. 1996: 97–104; Uganda Bureau of Statistics 2001: 97–108; 2007: 109–118; 2012: 98–104; National Council for Population and Development *et al.* 1989: 55–72; 1994: 83–92; Kichamu 1999; Muindi, Bicego 1999; Bicego, Kichamu 1999; Otieno, Omolo 2004; Kenya National Bureau of Statistics 2010: 103–112.

**Fig. 4.**  Population Growth Rates (annual %) in North and East Africa (1960–2011)

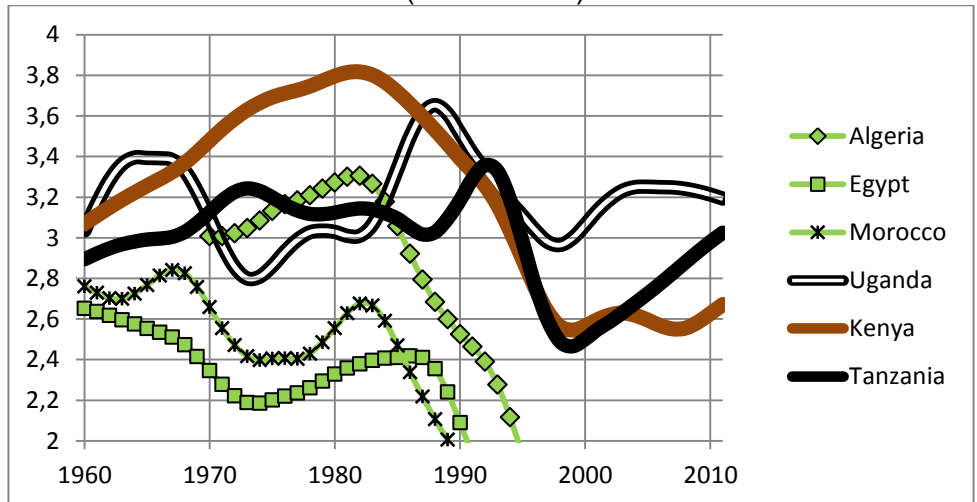

*Data source*: World Bank 2013: SP.POP.GROW.

It appears important to note that the escape from the Malthusian Trap was accompanied in North Africa by a fast increase of the productivity of labor in agriculture (see Fig. 5):

**Fig. 5.**  Dynamics of Labor Productivity in Agriculture (constant 2000 dollars per worker) in North and East Africa (1961–2011)

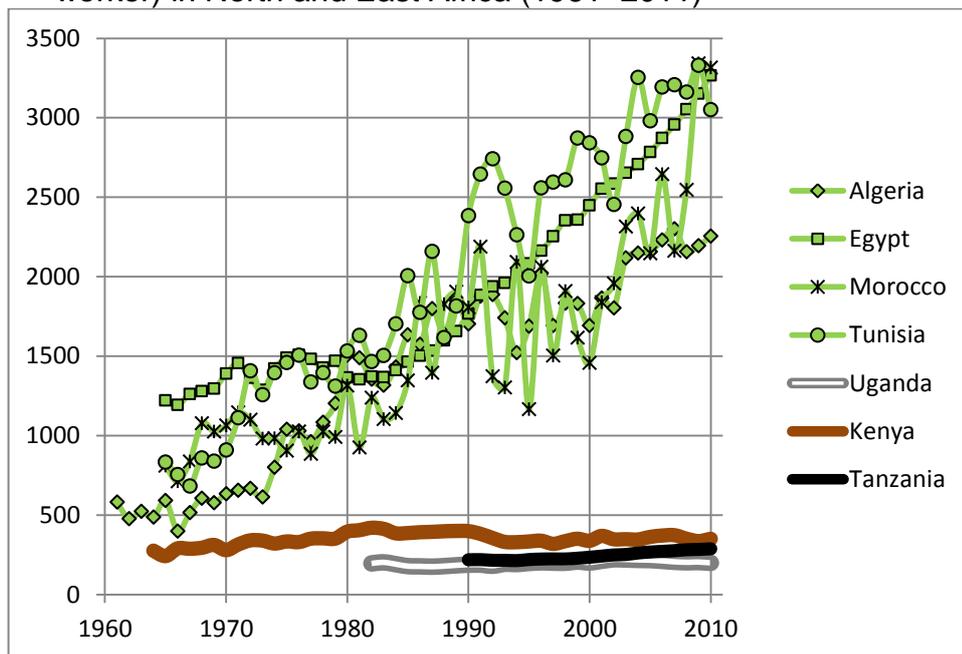

*Data source*: World Bank 2013: EA.PRD.AGRI.KD.

Quite predictably this was also accompanied by a sharp decline of percentage of economically active people employed in agriculture and now it constitutes 28.2% in Egypt, 17.7% in Tunisia, and 11.7% in Algeria (World Bank 2013: SL.AGR.EMPL.ZS)[8].

East Africa stands here in a sharp contrast. The most clear Malthusian economic-demographic dynamics is observed in the recent decades in Kenya (see Fig. 6):

---
[8] There is no direct data on this variable for North African countries in the early 1960s, but one may take into account the point that in 1960 the rural population constituted about two thirds of the overall population in those countries (UN Population Division 2012).

**Fig. 6.** Relative Dynamics of Population, GDP, and GDP per capita in Kenya, 1980–2009 (100 = 1980 level)

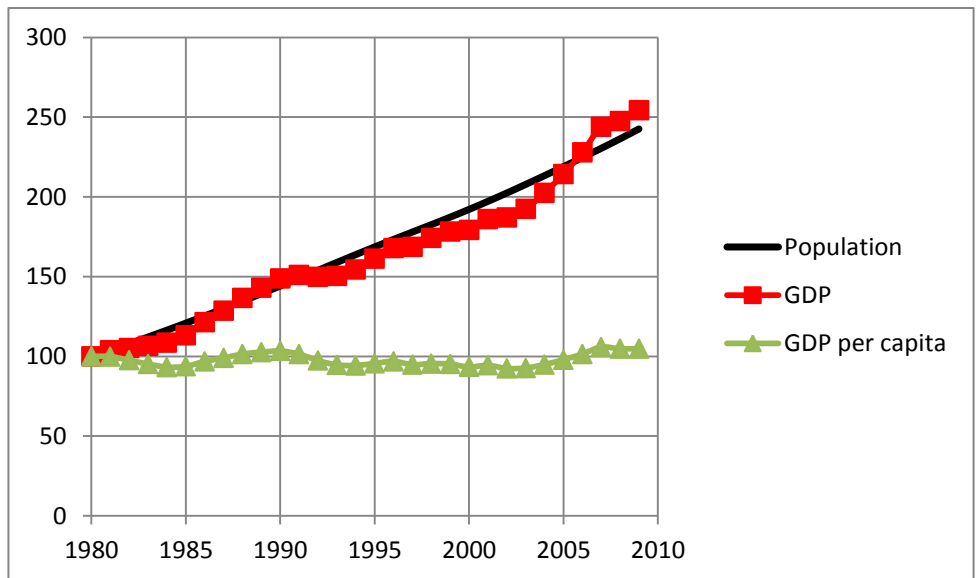

*Data source*: World Bank 2013: SP.POP.TOTL, NY.GDP.MKTP.PP.KD.

As we see, between 1980 and 2009 the Kenyan economy grew rather substantially, by about 150%. But the Kenyan population in those years grew as substantially – and, as a result, per capita GDP in Kenya in 2009 was almost the same as in 1980.

However, the situation in Uganda and Tanzania does not appear as simple. In both countries in recent years one could observe a rather substantial growth of per capita GDP (see Fig. 7):

**Fig. 7.** GDP per capita (PPP, constant 2005 international $) dynamics in Tanzania and Uganda, 1994–2011

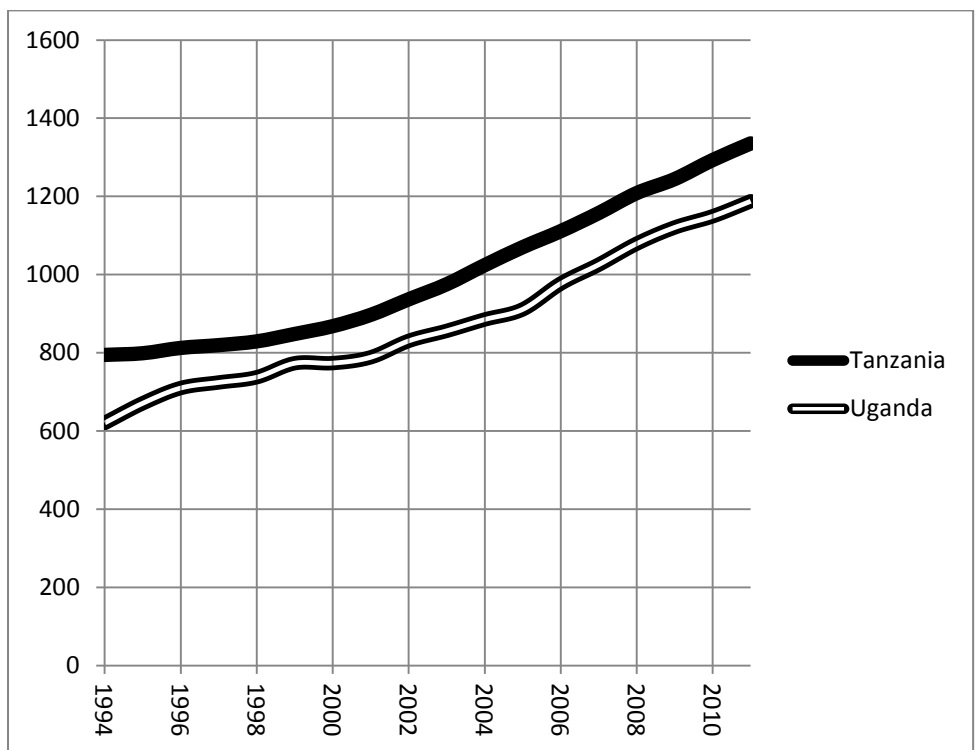

*Data source*: World Bank 2013: NY.GDP.PCAP.PP.KD

Why was not this converted in the growth of the per capita food consumption? It appears very important to answer this question, as the positive values of per capita GDP growth rates observed recently in most countries of Tropical Africa have made some experts believe that "(almost) the whole World – including most IF NOT ALL of Africa – shares features of Malthus' vision for escaping the trap he outlined" (Kenny 2010: 192, our emphasis).

To answer this question it appears necessary to consider the recent economic growth in East Africa by sector. Let us start with the dynamics of production of services in Uganda (see Fig. 8):

**Fig. 8.** Relative Dynamics of Population, Services Production, and Services Production per capita in Uganda, 1994–2011 (100 = 1994 level)

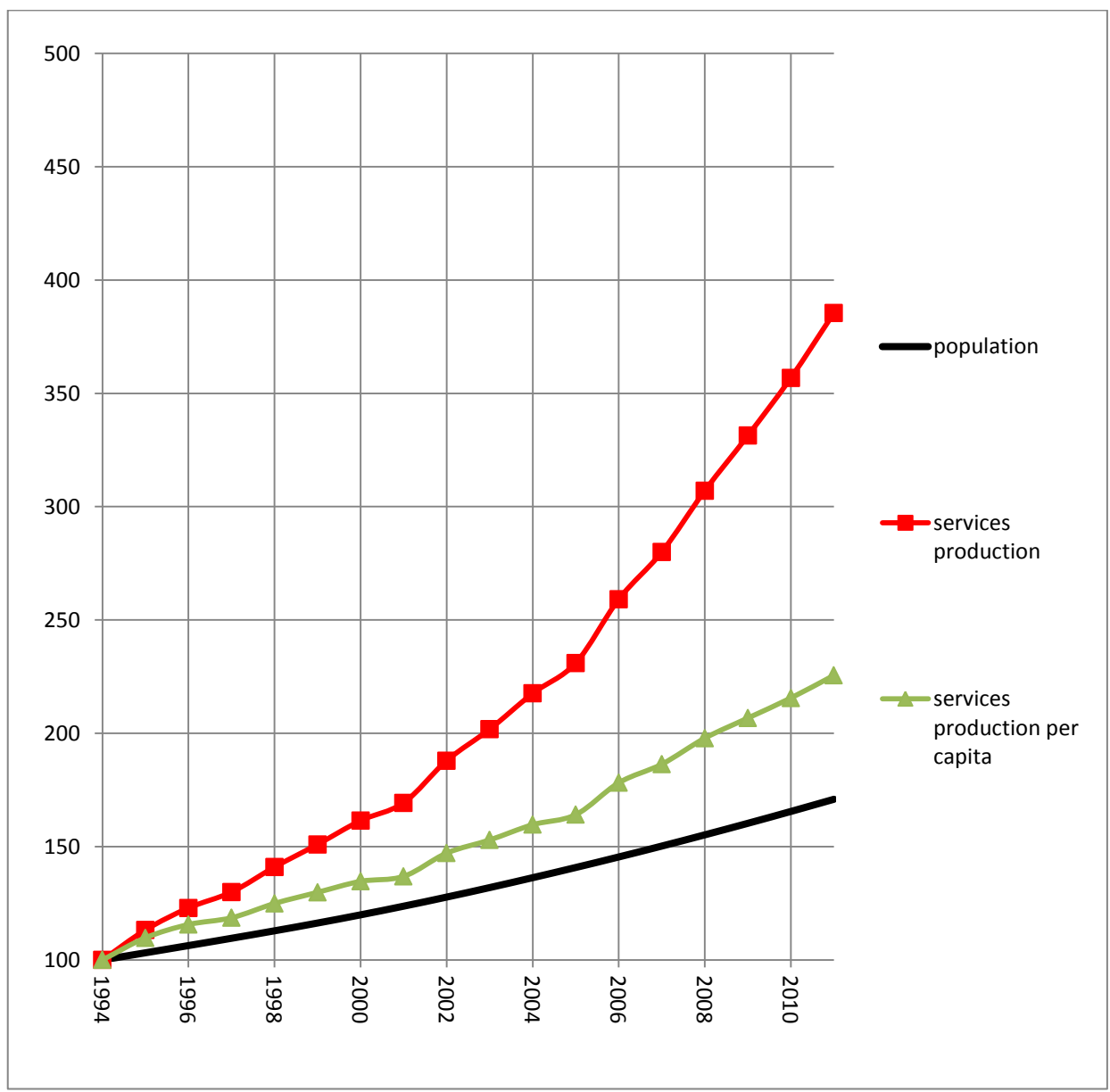

*Data source*: World Bank 2013: NV.SRV.TETC.KD, SP.POP.TOTL; the services production index was calculated on the basis of the World Development Indicator NV.SRV.TETC.KD: "Services, value added (constant 2000 US$)".

As we see, in these years Ugandan successes in this sector of its economy were really spectacular. Between 1994 and 2011 Ugandan population almost doubled. But growth rates in the service sector by far outpaced the population growth rate – production in this sector quadrupled, and per capita production of services increased more than twice.

Successes of industrial development in Uganda in these years were even more spectacular – between 1994 and 2011 industrial output grew in this country five times (see Fig. 9):

**Fig. 9.** Relative Dynamics of Population, Industrial Production, and Industrial Production per capita in Uganda, 1994–2011 (100 = 1994 level)

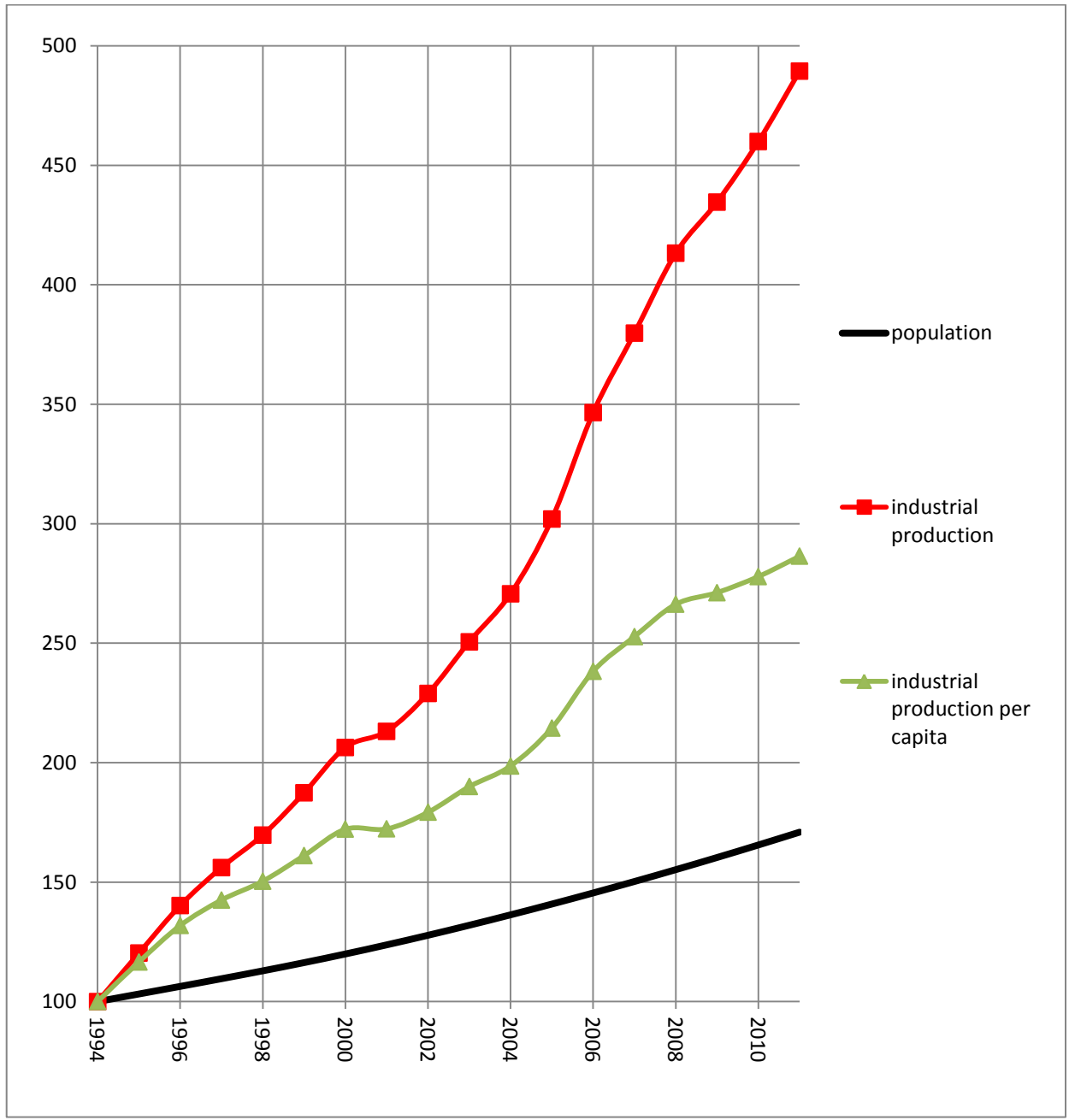

*Data source*: World Bank 2013: NV.IND.TOTL.KD, SP.POP.TOTL; the industrial production index was calculated on the basis of the World Development Indicator NV.IND.TOTL.KD: "Industry, value added (constant 2000 US$)".

But what about the sector where the majority of the Ugandan working population is still employed? Unfortunately, the situation in this sector is strikingly different from the success stories of the Ugandan non-agricultural enterprises (see Fig. 10):

**Fig. 10.** Relative Dynamics of Population, Agricultural Production, and Agricultural Production per capita in Uganda, 1994–2011 (100 = 1994 level)

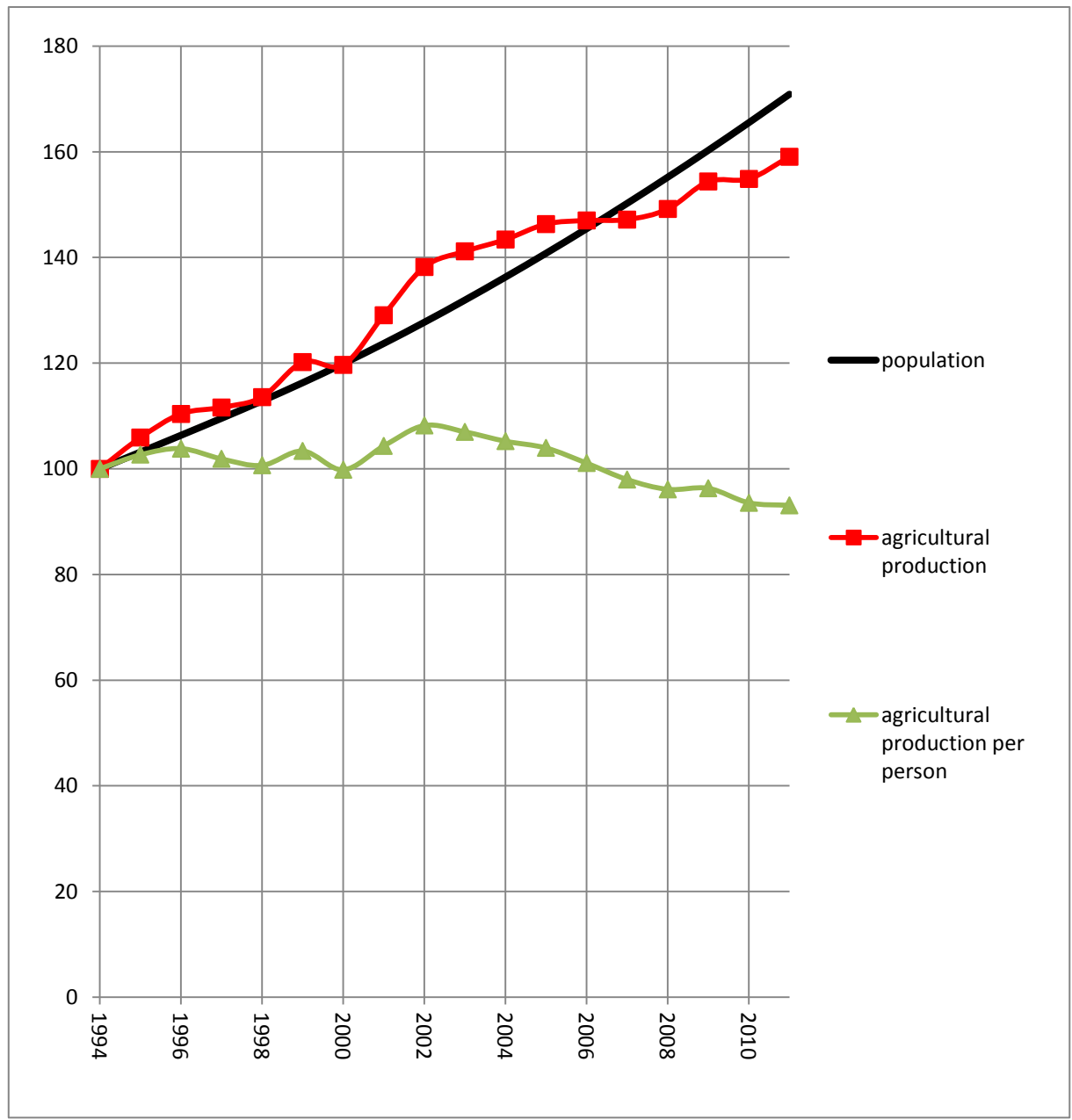

*Data source*: World Bank 2013: NV.AGR.TOTL.KD, SP.POP.TOTL; the agricultural production index was calculated on the basis of the World Development Indicator NV.AGR.TOTL.KD: "Agriculture, value added (constant 2000 US$)".

As we see, between 1994 and 2011 the growth of agricultural production in Uganda was still quite substantial (60%). But the population growth in Uganda in those years was even higher. And as a result the agricultural production per capita in Uganda declined. Note that this correlates rather well with the data on almost perfect stagnation of the agricultural labor productivity in Uganda in the recent years[9] (see above Fig. 5).

Actually one of the most important differences between contemporary North and East Africa is that in North Africa a rather small proportion of population is employed in agriculture, whereas in East Africa this is the overwhelming majority of the population (see Fig. 11):

---

[9] For some peculiarities of Tanzanian case see Appendix 1. On the phenomenon of the agricultural growth rate lagging behind the overall economic growth rate in Uganda see, e.g., Deininger, Okidi 2001: 124–125.

**Fig. 11.** Employment in agriculture, %, in North and East Africa in the 2000s

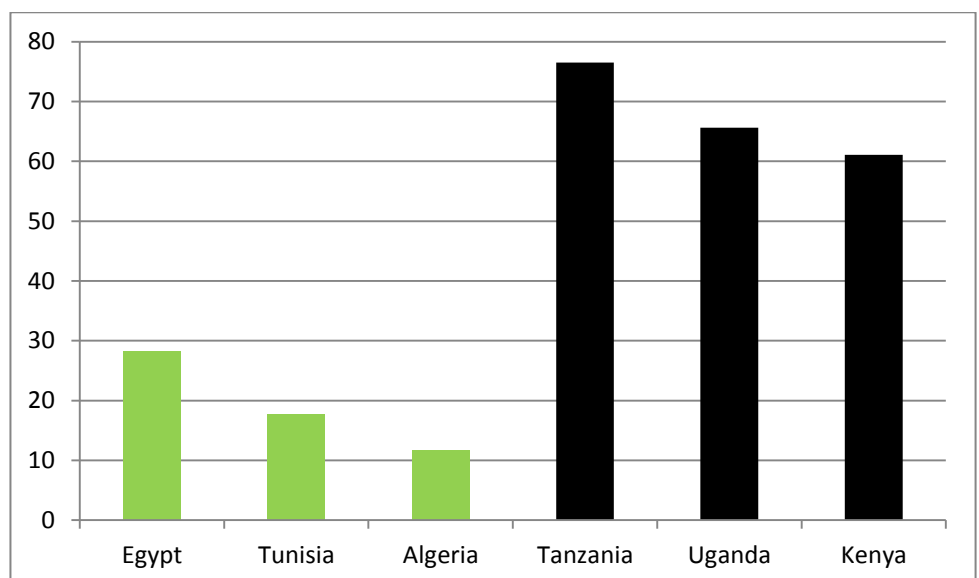

*Data source*: World Bank 2013: SL.AGR.EMPL.ZS.

Hence, the majority of the population of the East African countries will remain in the Malthusian Trap even against the background of an overall growth of GDP per capita if this growth is achieved through the productivity growth in non-agricultural sectors only with the agricultural labor productivity of the majority of rural population stagnating or declining.

Hence, a sustainable escape from the Malthusian trap can hardly be achieved in East Africa without a substantial increase in the productivity of labor of the majority of agricultural population. However, one of the main obstacles for this is just the rural overpopulation effectively blocking such increases.

As Roth and Fratkin (2005: 6) note, in East Africa "rapid population growth has affected rural and urban areas, where farmers increasingly move onto less productive lands to raise their crops and families". Population pressure leads to the reduction of the fallow periods[10], as a result of which the soil fertility fails to restore to original levels, as a result frequently various innovations (like fertilizers) just allow to prevent (at least partly) the decline of the current low agricultural labor productivity instead of increasing it (see, e.g., Bigsten, Kayizzi-Mugarewa 1999: 80).

Hence, the countries of East Africa will hardly be able to achieve a sustainable escape from the Malthusian Trap before they achieve a serious fertility decline.

However, why could not East Africa follow the "North African" path – that is, achieving the escape from the Malthusian Trap first, and accomplishing the fertility decline second?

One point is that North Africa escaped from the Malthusian Trap just 1–2 decades after the population growth rates exceeded 2.5% per year levels (implying the population doubling just within 30 years), whereas in East Africa such a growth continues for many decades and with any new decade the escape from the Malthusian Trap before the fertility decline becomes less and less likely.

---

[10] Those analysts who deny the presence of overpopulation problems in Tropical Africa love to say things like "only 12% of all potential arable land is under cultivation [in Sub Saharan Africa]" (Hayes 2012: 114); however, they tend not to take into consideration the point that "new" land being brought under cultivation in Tropical Africa under population pressure in most cases is not virgin, but it is rather fallow land whose straightforward cultivation leads to the reduction of fallow periods, inadequate soil fertility restoration, soil degradation, and decrease of labor productivity, whereas against such a background serious efforts should be taken even in order just to retain the agricultural labor productivity at old level. E.g., "Blackie et al. (1998) observed that because of increasing population density and declining land availability 'maize is now grown in continuous cultivation rather than as part of a fallow [rotation] which traditionally used to restore soil fertility and reduce the build up of pests and diseases. The soil resource base is now being degraded with a consequent reduction of yield' " (Barahona, Cromwell 2005: 159).

But, if there are such successes in urban sectors, why just not to move the majority of population from stagnant villages to vibrant cities?

To answer this question it makes sense to consider the urban population dynamics in North and East Africa (see Fig. 12):

**Fig. 12.** Percentage of Urban Population in North and East Africa, 1970–2010[11]

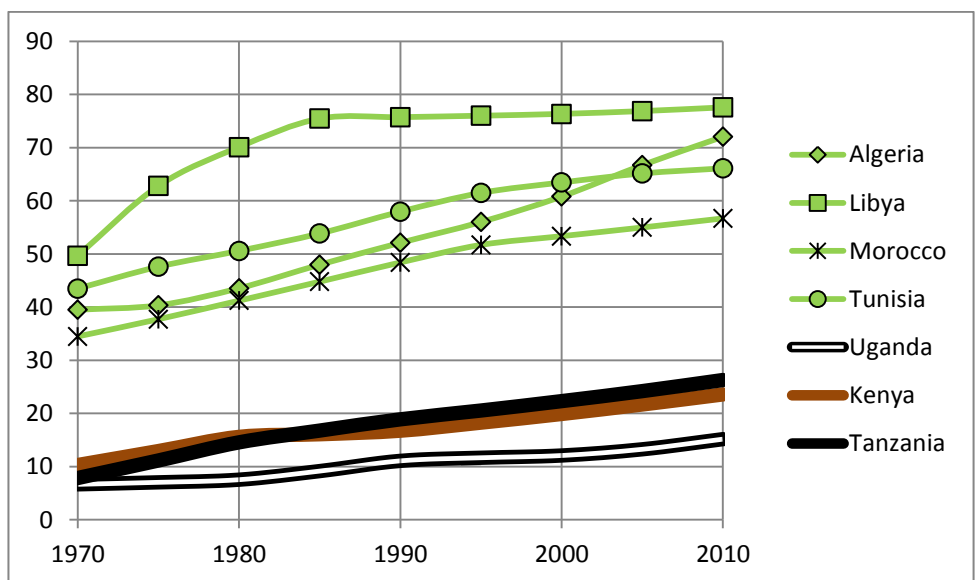

*Data source*: UN Population Division 2012.

As we see, even now the proportion of urban population in East Africa is substantially lower than it was in North Africa at the start of its escape from the Malthusian Trap. And if the recent urban population percentage growth rate, say, in Uganda continues, this country will only reach the 1970 level of the least urbanized North African country, Morocco, in the late 2050s.

It appears also relevant to consider the recent UN population division medium projections for the three main countries of East Africa. Note that those projections are based on the assumption that the fertility decline in East Africa will accelerate in the forthcoming decades (see Fig. 13):

---

[11] This diagram does not display the curve for Egypt, as the official Egyptian data on the variable in question (reproduced by the *World Urbanization Prospects*) appear irrelevant due to the huge underestimation of the actual numbers of urban population by the official Egyptian statistics in the recent years: "Though according to government statistics Egypt remains a largely rural country, many villages have expanded, some to over 100,000 inhabitants, but have not been reclassified as towns… The population of a medium-sized Egyptian 'village' [is in the range] 10,000 to 10,500 people; naturally, as there are villages with much less population, there invariably exist villages with much greater populations. According to global standards, these 'villages' with more than 10,000 dwellers ought to be classified as small towns (which would account for a prevalence of non-agricultural employment [in these settlements]), but… in this respect Egyptian statistics failed to keep pace with the rate of village population growth, and reclassification [of those 'villages' into 'towns'] is yet to be expected" (Zinkina, Korotayev 2013: 34–35).

**Fig. 13.** UN Population Division projections of TFR (births per woman) dynamics in East Africa underlying their medium population forecast for the main East African countries

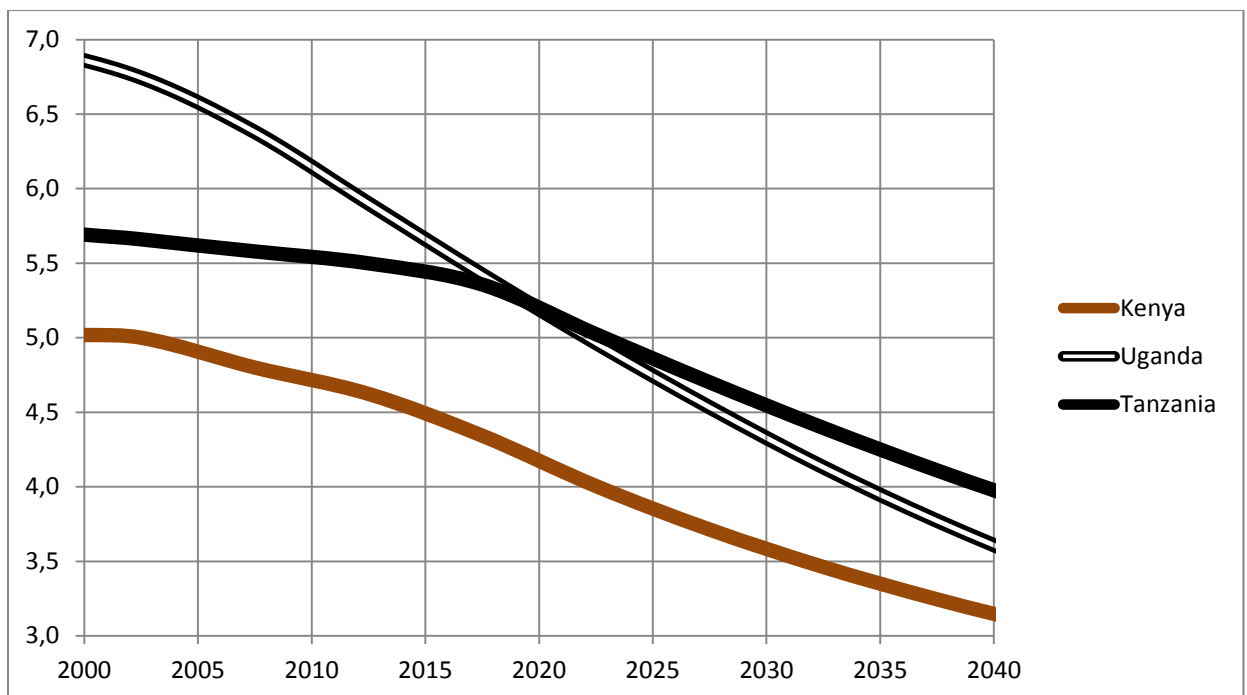

*Data source*: UN Population Division 2013.

Still these projections forecast the following population dynamics for East Africa (see Fig. 14):

**Fig. 14.** UN Population Division medium population forecast for the main East African countries, millions

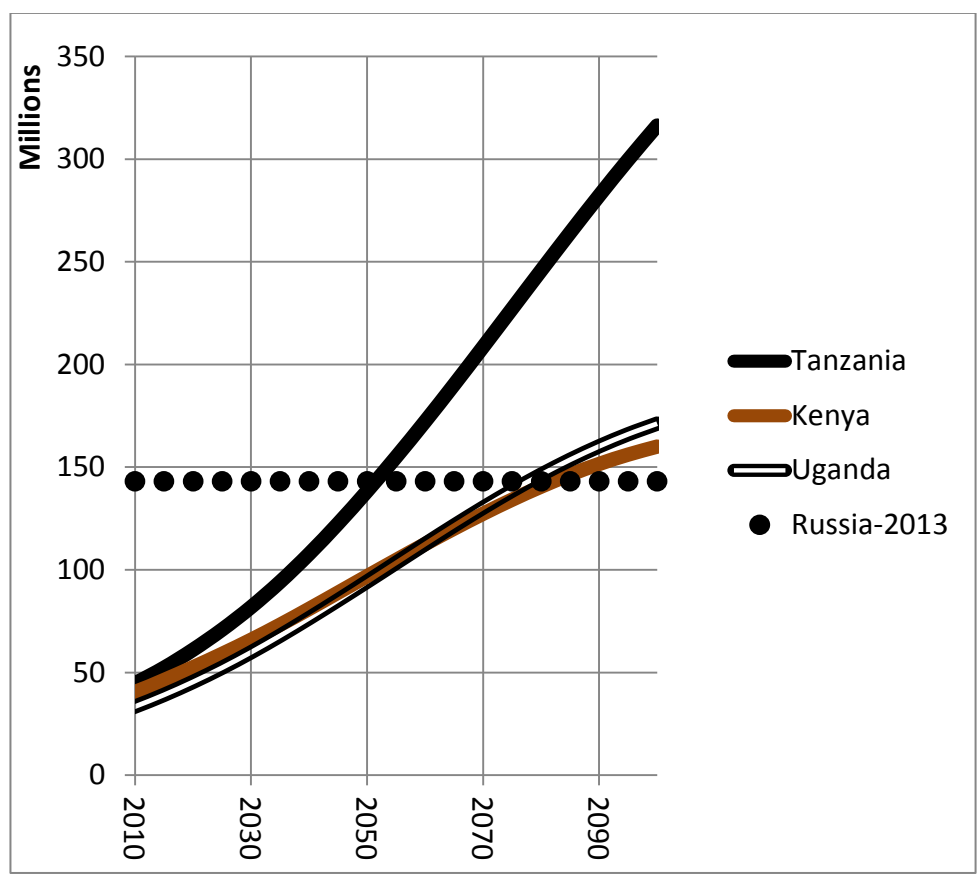

*Note*: black dotted line presents the current population of Russia for comparison. *Data source*: UN Population Division 2013.

As we see, even with the acceleration of fertility decline forecasted by the UN Population Division, the population of Kenya and Uganda will exceed the population of Russia in the second half of the century. Tanzania will reach Russia in terms of population by 2050 and is projected to have twice the Russian population by 2100. And Tanzania is almost bound to have population of Russia, as, according to the estimates of the UN Population Division (2013), the number of children below 5 in Tanzania is already almost the same as in Russia.

Note that without more substantial fertility declines (which would imply the introduction of compulsory universal secondary education, serious family planning programs of the Rwandan type [see, *e.g.*, Kinzer 2007; WHO 2008; Lu *et al.* 2012]), and the rise of legal age of marriage with parental consent) East Africa is likely to face two options, if it would like to escape the Malthusian Trap:

1) to move no less than 50% of its rural populations to dynamically growing urban sectors (that is, to increase the urban population proportion growth rate substantially in comparison with what was observed in recent years [see Fig. 12]). Note that, in conjunction with the UN Population Division projections this would imply the growth of the urban population if East Africa by about 80 million people by 2050 and by c. 160 million people by the end of this century. This would be tantamount to the creation in a few forthcoming decades in East Africa of about 80 new cities of the size of Koeln. Most likely such a scenario would rather produce an explosive growth of slums, urban overpopulation, social explosions and so on (cf. Korotayev *et al.* 2011).

2) On the other hand, if in East Africa the current trend of the urban population percentage growth continues, this would imply that the rural population of East Africa (Tanzania + Kenya + Uganda) would be five timesconsiderable fertility decline. Hence, it appears that it has lost the moment when it could follow the "North African" model of the escape from the Malthusian Trap. Note also that in the most important country of North Africa, in Egypt, the escape from the Malthusian Trap did not lead to fertility decline automatically, but it demanded a rather considerable family planning effort.

Mubarak's administration was well aware of the threat hidden in the growing gap between declining death rate and stably high birth rate, and almost since the beginning of Mubarak's reign (1981) it started taking measures aimed at bringing down the birth rate (see, *e.g.*, Fargues 1997: 117–118). However, only in the second half of the 1980s the government managed to develop a really efficient program of such measures. This program was performed by the Egyptian government in collaboration with USAID program aimed at wide-scale introduction and distribution of family planning (Moreland 2006). Religious leaders (from al-Azhar *sheikh*s to local *imam*s) were involved in the program to disseminate (in their *fatwah*s and sermons) the idea that family planning was not adverse to *al-Qur'an*; on the contrary, it is good, as having less children makes it easier for the parents to give them a happy childhood and good education (Ali 1997). This strategy proved essentially effective, as during 5 years (1988–1992) total fertility rate in Egypt fell from 5 to 4 children per woman.

One may wonder if East Africa should not rather follow now the "Bangladeshi scenario". In Bangladesh the fertility decline PRECEEDED the start of successful development.

Successful development only started when in the 1990s Bangladeshi TFR fell below 4 children per woman level (due to implementation of rather effective family planning programs [see, e.g., ). Before that Bangladesh rather followed the "East African path" – very fast population growth "ate" almost all the (rather substantial) GDP growth that took place in Bangladesh. Note that though Bangladeshi GDP grew between 1970 and 1995 (i.e. in 25 year period) more than twice; however, the per capita GDP remained almost the same. By contrast, after Bangladesh had managed to bring TFR below 4, the per capita GDP in this country increased in 15 years by about 100% (see Figs. 15 and 16):

**Fig. 15.** TFR Dynamics in Bangladesh (1960–2010)

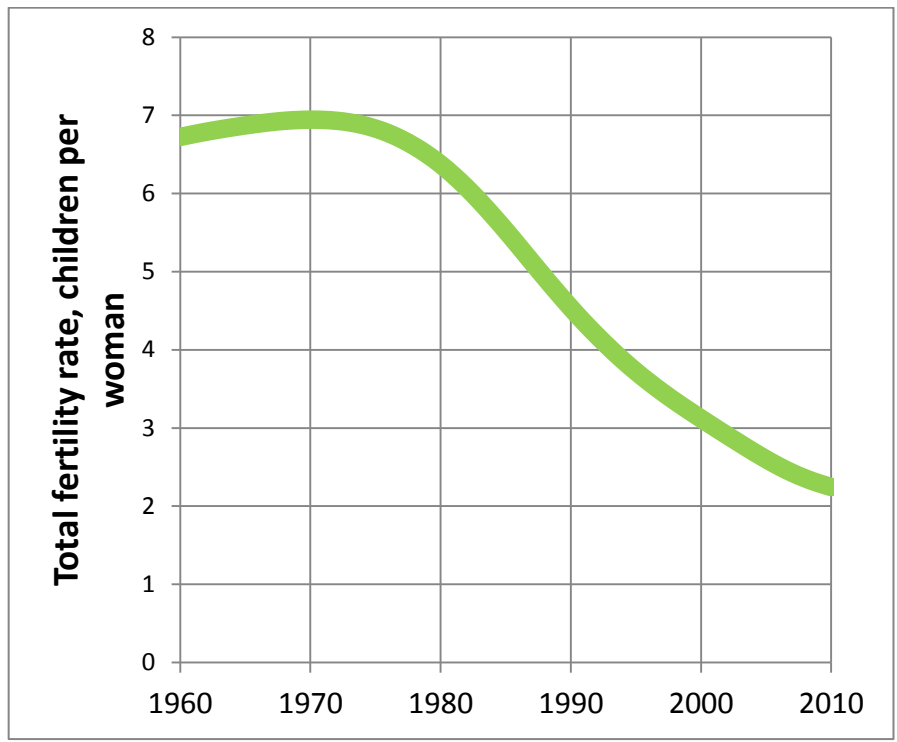

*Data source*: World Bank 2013: SP.DYN.TFRT.IN.

**Fig. 16.**   Relative Dynamics of Population, GDP, and GDP per capita in Bangladesh, 1970–2008 (100 = 1970 level)

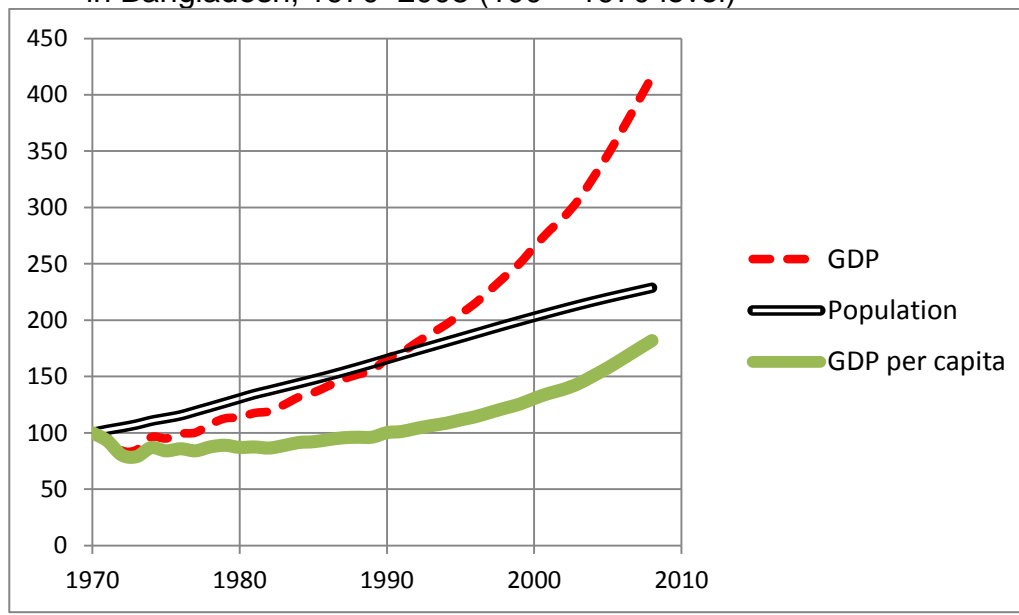

*Data source*: World Bank 2013: SP.POP.TOTL, NY.GDP.MKTP.PP.KD.

Still there was an appreciable (albeit very slow) growth of per capita GDP in Bangladesh in 1980–1995. However, this growth failed to be translated into a significant improvement of the life of the majority of the Bangladeshi population, which is evidenced by the dynamics of average food consumption that remained during this period at a very low level that is typical for Tropical African countries trapped in the Malthusian trap and did not exhibit any tendency to increase (that is also rather characteristic for this type of countries) (see Fig. 17):

**Fig. 17.** Average per capita food consumption dynamics (kcal per capita per day) in Bangladesh, Kenya, Tanzania, and Uganda (1980–1995)

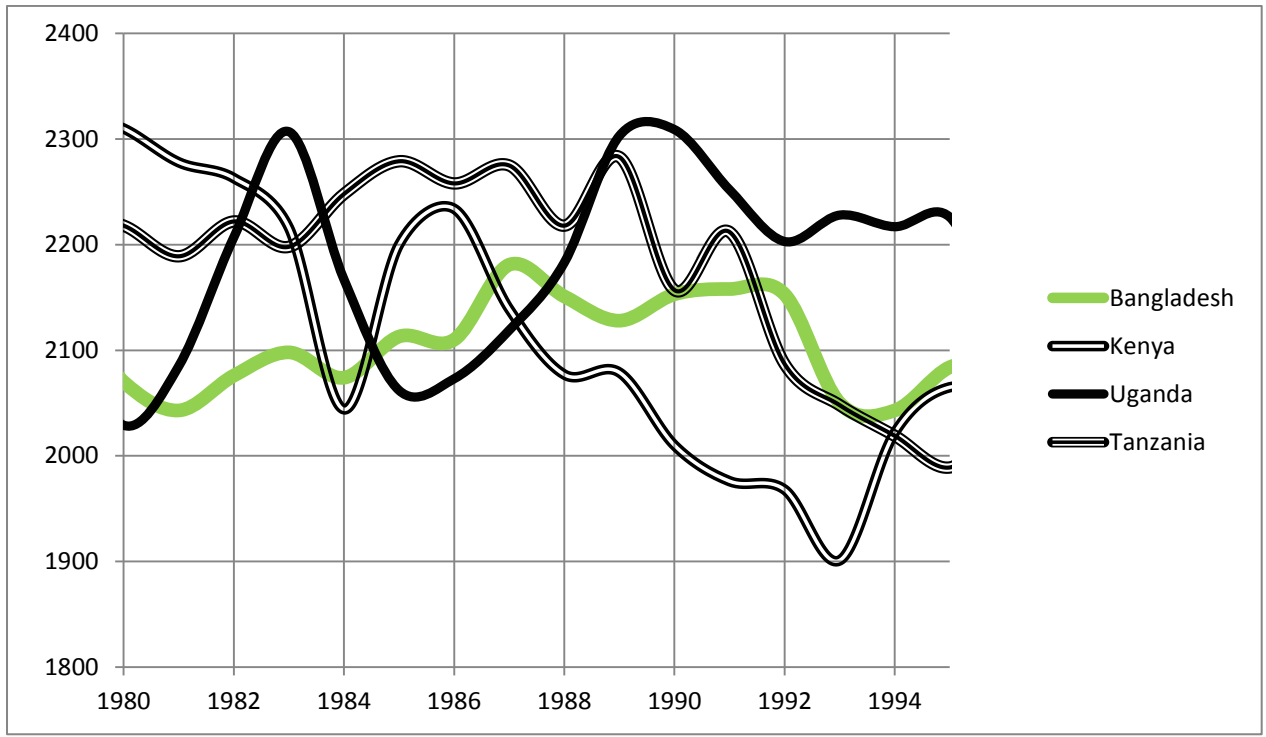

*Data source*: FAO 2013.

Thus, when fertility in Bangldeshi remained close to Tropical African levels (see Fig. 18 below), average per capita food consumption levels also remained extremely close to Tropical African ones (see Fig. 17 above).

**Fig. 18.** TFR dynamics in Bangladesh, Kenya, Iran, Uganda, and Tazania (1980-1995)

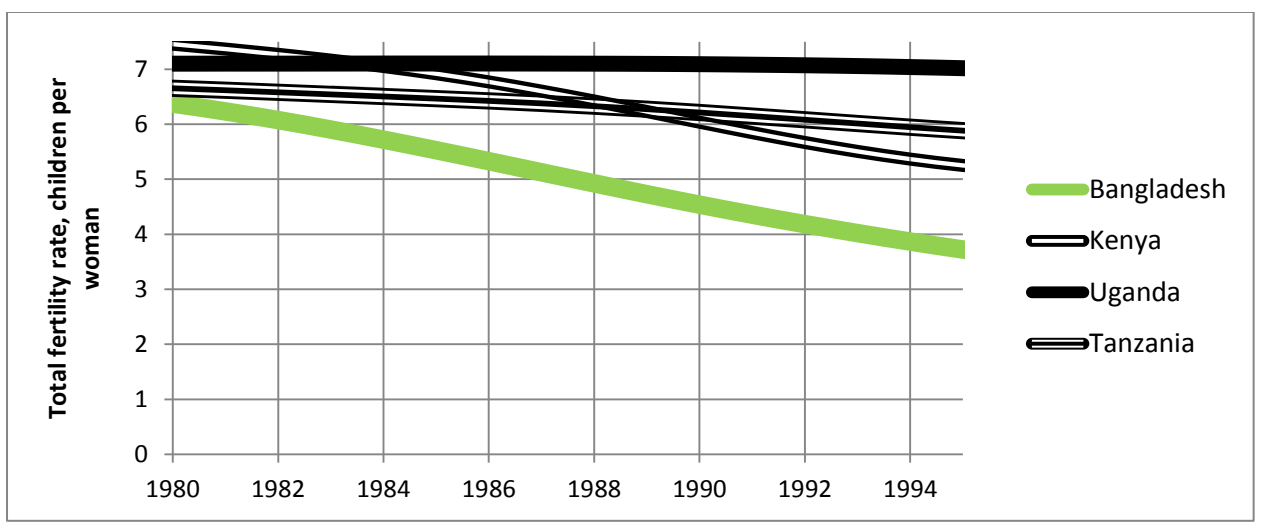

*Data source*: World Bank 2013: SP.DYN.TFRT.IN.

But why did in 1980–1995 the average per capita food consumption in Bangladesh stall against the background of quite visible growth of per capita GDP in those years. The point is that this growth was achieved through the increase in productivity in non-agricultural sectors of

economy – in industry, construction, transportation, trade, financial sector, and so on. However, one should take into account that at that time a rather small minority of Bangladeshi population lived in the cities (the same was true for the East African countries as well [see Fig. 19]) – note that the overwhelming majority of population is still rural both in Bangladesh and in Tropical African countries in question (see Fig. 20):

*Fig. 19.* **Urban population (% of total) in Bangladesh, Tanzania, and Kenya in 1995**

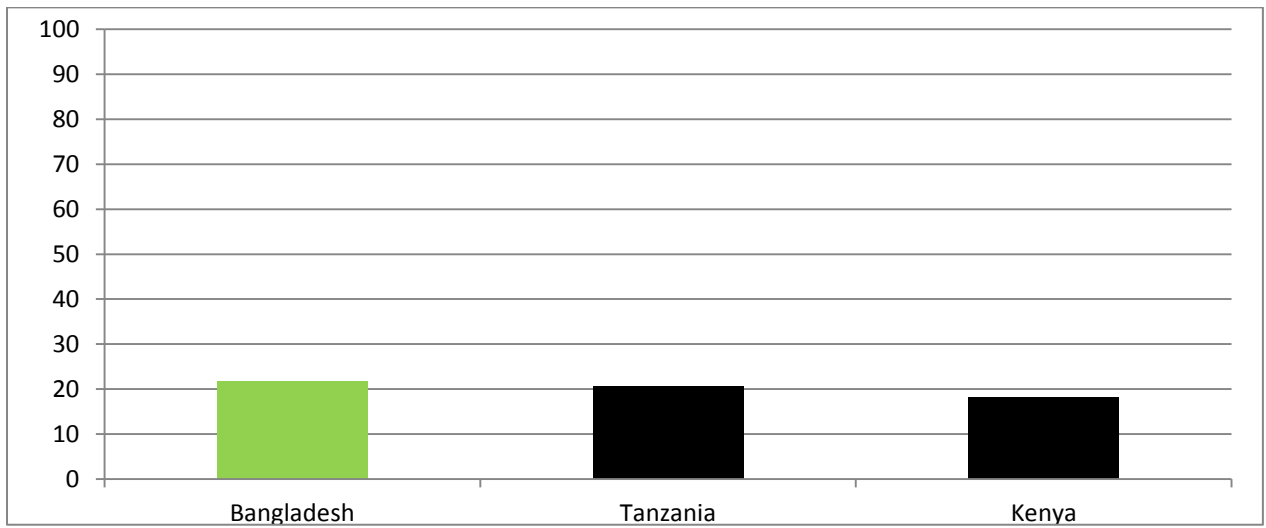

*Data source*: UN Population Division 2012.

Fig. 20. **Dynamics of urban population (% of total) in Bangladesh and East Africa, 1960-2010**

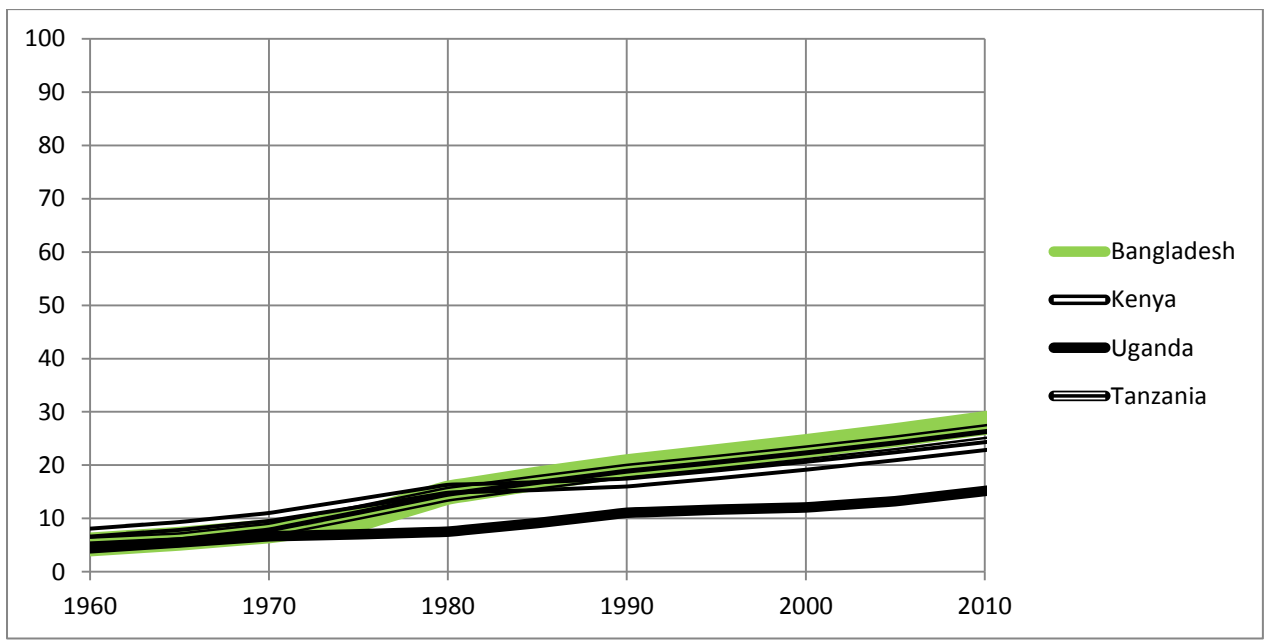

*Data source*: UN Population Division 2012.

Against this background it is highly important to notice that in 1980–1995 one could observe a stall of the productivity of labor in agriculture – it was not only very low (quite comparable with its contemporary levels in Tropical African countries still trapped in the Malthusian trap); what is more it hardly grew during the period in question, and it is quite clear that the fact that the growth of productivity of agricultural labor during this period was blocked by explosive rural overpopulation was the main cause of the stall in the growth of the average per capita food consumption (see Fig. 21):

Fig. 21. **Labor productivity in agriculture (value added per worker per year, constant 2000 dollars) in Bangladesh and East Africa, 1980-1995**

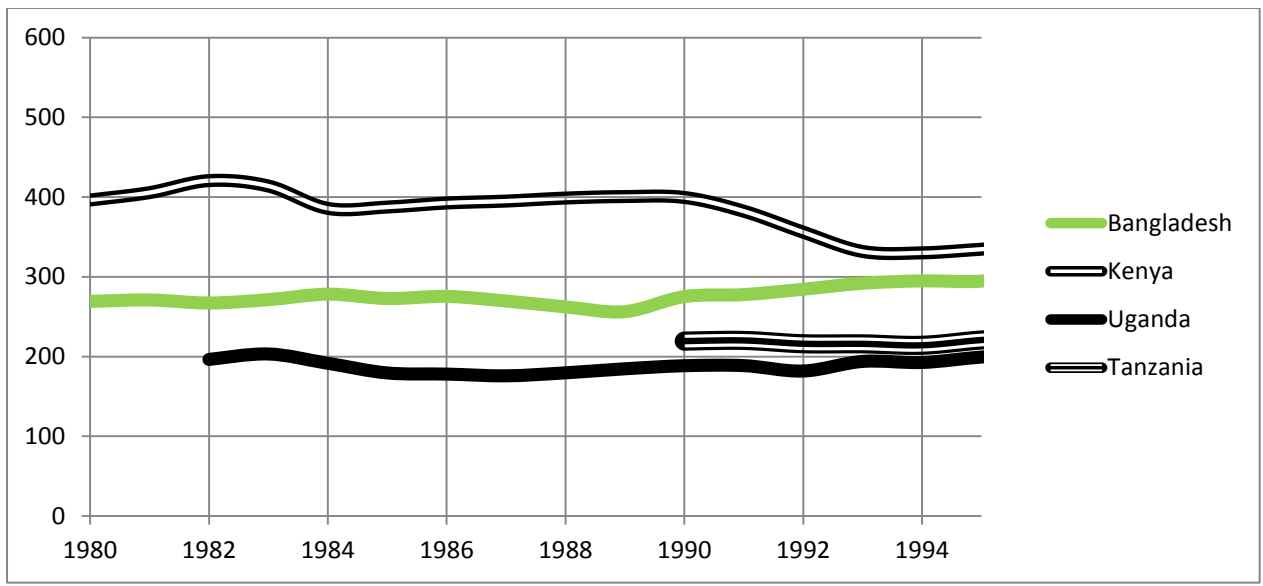

*Data source*: World Bank 2013: EA.PRD.AGRI.KD.

The overall picture of relative dynamics in Bangladesh in the period before it achieved substantial decline in fertility levels looks as follows (Fig. 22):

Fig. 22. **Dynamics of GDP per capita, agricultural labor productivity, and average per capita food concumption in Bangladesh (1980-1995, 100 = 1980 level)**

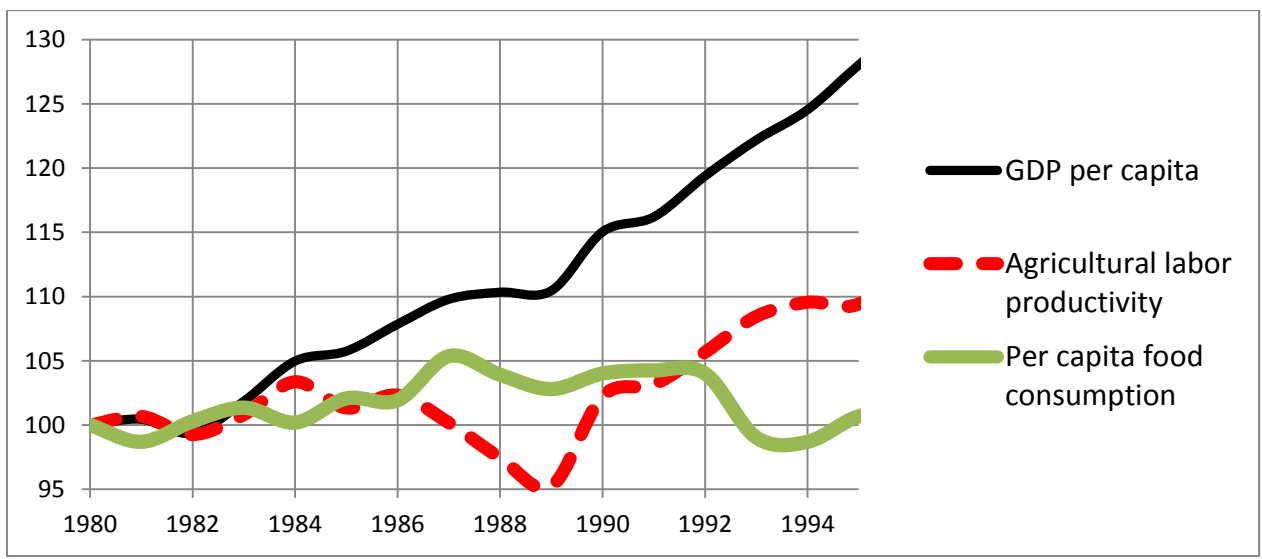

*Data source*: World Bank 2013: EA.PRD.AGRI.KD, SP.POP.TOTL, NY.GDP.MKTP.PP.KD; FAO 2013.

As we see against the background of huge population pressure even quite noticeable per capita GDP growth failed to get translated into a comparatively noticeable growth of agricultural labor productivity, which (within the context of the overwhelming majority of population still being rural) resulted in the absence of any noticeable improvement in the average per capita food consumption at all.

Note that in some East African countries such a dynamics is observed till now. For example, for Uganda it looks as follows (Fig. 23):

Fig. 23. **Dynamics of GDP per capita, agricultural labor productivity, and average per capita food concumption in Uganda (1990-2009, 100 = 1990 level)**

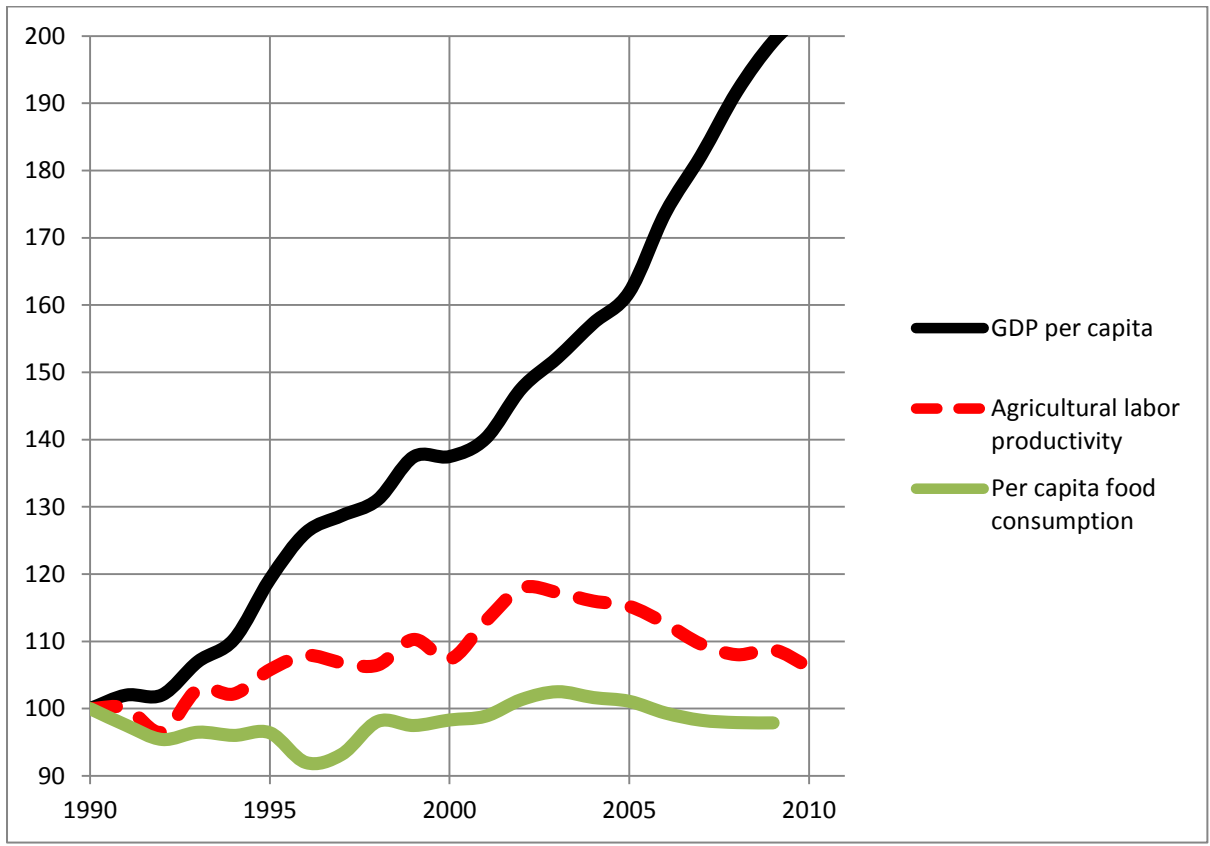

*Data source*: World Bank 2013: EA.PRD.AGRI.KD, SP.POP.TOTL, NY.GDP.MKTP.PP.KD; FAO 2013.

As we see the pattern which is observed till now in Uganda is almost identical with the one which had been observed in Bangladesh before this country managed to achieve a really substantial fertility decline – quite a noticeable per capita GDP growth fails to get translated into a comparatively noticeable growth of agricultural labor productivity, which (within the context of the overwhelming majority of population still being rural) results in the absence of any noticeable improvement in the average per capita food consumption at all.[12] This demonstrates again that we should not expect that the comprehensive development will "automatically" bring about all the necessary fertility declines (this belief seems to be analogous to the famous belief of some economists that the "invisible hand of market" will sort everything out automatically). The situation is just opposite in the countries trapped in the Malthusian trap (which is the case for the most of the present-day Tropical African countries) – in those countries the achievement of the substantial fertility decline is a precondition of successful comprehensive development.

---

[12] Note that this is not the worst possible scenario for the present-day Tropical Africa (at least in recent decades we do not observe there any pronounced trends toward decline of those variables. For example, in Kenya where after 1980 population grew as fast as GDP, which resulted in stagnation of per capita GDP and a long-term trend toward decline in agricultural labor productivity and average per capita food consumption.

Bangladesh managed to get in the trajectory of really successful comprehensive development only after by the mid-1990s it had managed to achieve a really substantial fertility decline (see Fig. 24):

*Fig. 24.* **Dynamics of agricultural labor productivity, and average per capita food concumption in Bangladesh (1980-2009, 100 = 1980 level)**

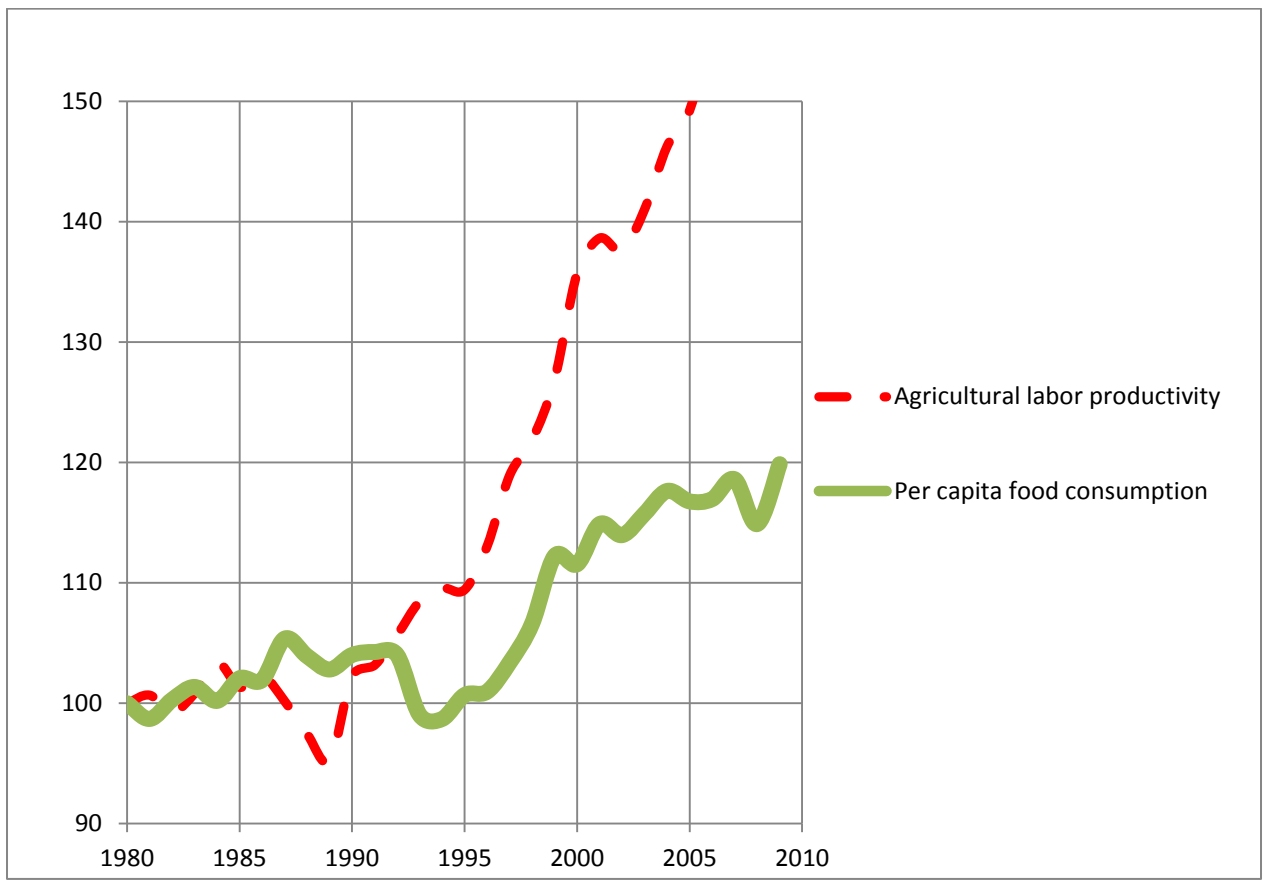

*Data source*: World Bank 2013: EA.PRD.AGRI.KD; FAO 2013.

Only after that Bangladesh started its way out of the Malthusian trap. Only after that it managed to outstrip its Tropical African analogues (that failed to achieve similar fertility declines and that, consequently, continue being trapped in the Malthusian trap) (see Figs. 25 and 26):

Fig. 25. **Labor productivity in agriculture (value added per worker per year, constant 2000 dollars) in Bangladesh and East Africa, 1995-2010**

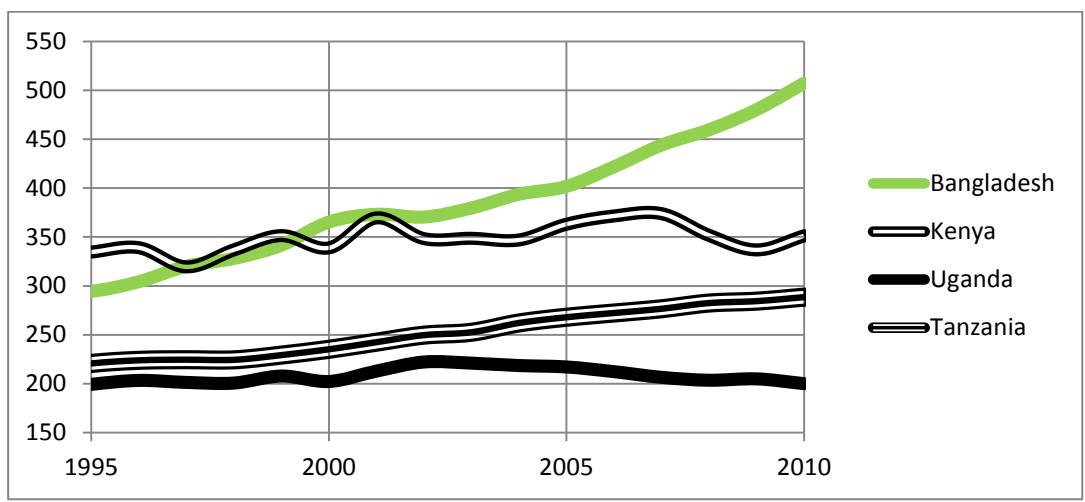

*Data source*: World Bank 2013: EA.PRD.AGRI.KD.

*Fig. 26.* **Average per capita food consumption dynamics (kcal per capita per day) in Bangladesh, Kenya, Tanzania, and Uganda 1990-2009**

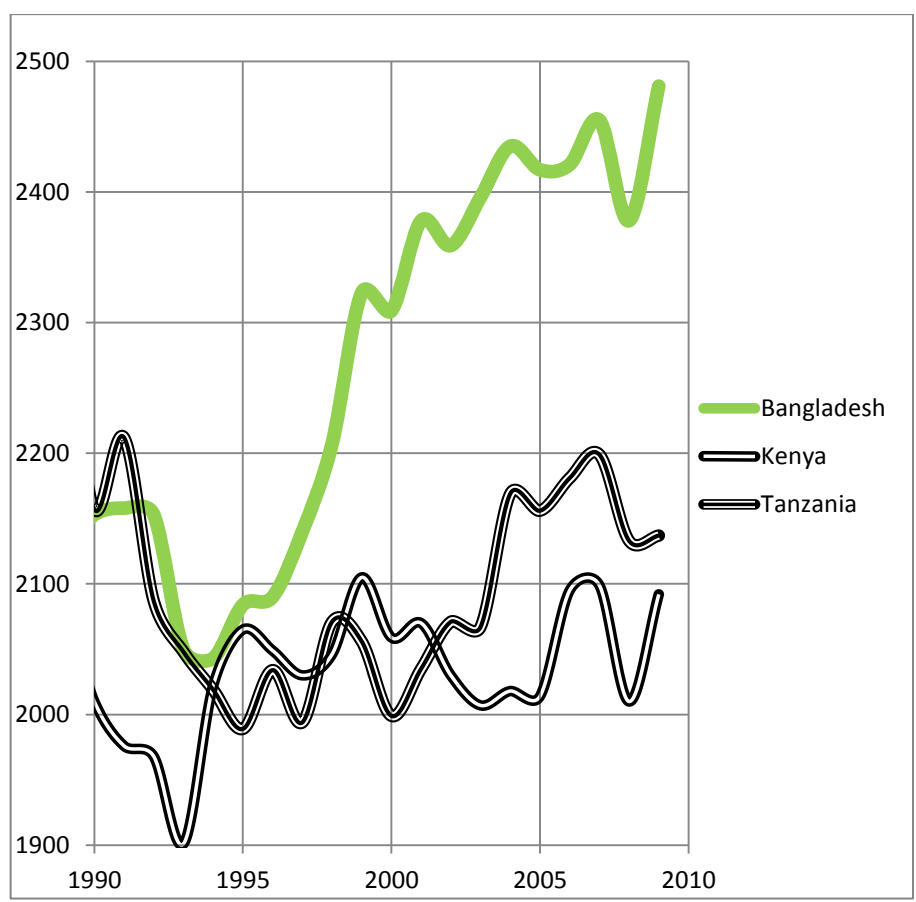

*Data source*: World Bank 2013: EA.PRD.AGRI.KD; FAO 2013.

Thus, the similarity of Bangladeshi pattern to the situation in modern East African countries is striking. This demonstrates again that we should not expect that the comprehensive development will "automatically" bring about all the necessary fertility declines. The situation is just opposite in the countries trapped in the Malthusian trap (which used to be the case for Bangladesh and still is the case for the most of the present-day Tropical African countries) – in those countries the achievement of substantial fertility decline is a precondition of successful comprehensive development.

**Conclusion**

Thus there are grounds to maintain:

1) that the main countries of East Africa (Uganda, Kenya, and Tanzania) have not escaped the Malthusian Trap yet;
2) that this countries are not likely to follow the "North African path" and to achieve this escape before they achieve serious successes in their fertility transition;
3) that East Africa is unlikely to achieve this escape if it does not follow the "Bangladeshi path" and does not achieve really substantial fertility declines in the foreseeable future, which would imply the introduction of compulsory universal secondary education, serious family planning programs of the Rwandan type, and the rise of legal age of marriage with parental consent. Such measures should of course be accompanied by the substantial increases in the agricultural labor productivity and the decline of the percentage of population employed in agriculture.

**Acknowledgement**

This article is an output of a research project implemented as part of the Basic Research Program at the National Research University Higher School of Economics (HSE).

**References**

Ali, K. A. (1997). Modernization and Family Planning Programs in Egypt. *Middle East Report, 205*, 40–44.

André, C., & Platteau, J.-P. (1998). Land relations under unbearable stress: Rwanda caught in the Malthusian trap. *Journal of Economic Behavior & Organization, 34*(1), 1–47.

Artzrouni, M., & Komlos, J. (1985). Population Growth through History and the Escape from the Malthusian Trap: A Homeostatic Simulation Model. *Genus, 41*(3–4), 21–39.

Badran, M., & Laher, I. (2011). Obesity in Arabic-Speaking Countries. *Journal of Obesity, 2011*, 1–9.

Barahona, C., & Cromwell, E. (2005). Starter Pack and Sustainable Agriculture. In S. Levy (Ed.), *Starter Packs: A Strategy to Fight Hunger in Developing Countries* (pp. 155–174). Cambridge, MA: CABI Publishing.

Bicego, G., & Kichamu, G. (1999). Adult and Maternal Mortality. In *Kenya Demographic and Health Survey 1998* (pp. 161–166). Calverton, MD: National Council for Population and Development (NCPD), Central Bureau of Statistics (CBS) (Office of the Vice President and Ministry of Planning and National Development) [Kenya], and Macro International Inc. (MI).

Bigsten, A., & Kayizzi-Mugerwa, S. (1999). *Crisis, Adjustment and Growth in Uganda. A Study of Adaptation in an African Economy.* London: Macmillan.

Bongaarts, J., & Sinding, S.W. (2009). A Response to Critics of Family Planning Programs. *International Perspectives on Sexual and Reproductive Health, 35*(1), 39–44.

Bureau of Statistics [Tanzania] & Macro International Inc. (1997). *Tanzania Demographic and Health Survey 1996.* Calverton, MD: Bureau of Statistics and Macro International.

Chu, C. Y. C., & Lee, R. D. (1994). Famine, Revolt, and the Dynastic Cycle: Population Dynamics in Historic China. *Journal of Population Economics, 7*, 351–378.

Clark, G. (2007). *A Farewell to Alms: A Brief Economic History of the World.* Princeton, NJ: Princeton University Press.

Conley, D., McCord, G., & Sachs, J. (2007). *Africa's Lagging Demographic Transition: Evidence from Exogenous Impacts and Agricultural Technology*. Cambridge, MA: National Bureau of Economic Research.

Deininger, K., & Okidi, J. (2001). Rural Households: Incomes, Productivity, and Nonfarm Enterprises. In R. Reinikka & P. Collier (Eds.). *Uganda's Recovery. The Role of Farms, Firms, and Government* (pp. 123–176). Washington, DC: The World Bank.

Egypt Ministry of Health, National Population Council, El-Zanaty and Associates, && ORC Macro. 2009. *Egypt Demographic and Health Survey (EDHS) – 2008*. Cairo: Egypt Ministry of Health.

FAO (Food and Agriculture Organization of the United Nations). (2009). *The State of Food Insecurity in the World. Economic crises – impacts and lessons learned.* Rome: Food and Agriculture Organization of the United Nations.

FAO (Food and Agriculture Organization of the United Nations). (2014). *FAOSTAT. Food and Agriculture Organization Statistics.* URL: http://faostat.fao.org/site/609/default. aspx#ancor.

Fargues, P. (1997). State Policies and the Birth Rate in Egypt: From Socialism to Liberalism. *Population and Development Review, 23*(1), 115–138.

Galor, O., & Weil, D. (1999). From Malthusian Stagnation to Modern Growth. *American Economic Review, 89*(2), 150–154.

Hayes, N. (2012). Population and Poverty in Sub Saharan Africa. In M. Kennet (Ed.), *Green Economics: Voices of Africa* (pp. 103–116.). Tidmarsh: The Green Economics Institute.

Joint FAO/WHO/UNU Expert Consultation. (1985). *Energy and protein requirements*. Geneva: World Health Organization (World Health Organization Technical Report Series 724).

Kaijuka, E. M., Kaija, E. Z. A., Cross, A. R., & Loaiza, E. (1989). *Uganda Demographic and Health Survey 1988/1989.* Entebbe: Ministry of Health.

Kenny C. 2010. Is Anywhere Stuck in a Malthusian Trap? *Kyklos, 63*(2), 192–205.

Kenya National Bureau of Statistics (KNBS) & ICF Macro. 2010. *Kenya Demographic and Health Survey 2008-09*. Calverton, MD: KNBS and ICF Macro.

van Kessel-Hagesteijn, R. (2009). Dynamics in political centralization processes – the various faces of 'decline*'. International Symposium on Early State Formation Handbook* (pp. 46–63). Peking: National Academy of China.

Kichamu, G. (1999). Early Childhood Mortality. *Kenya Demographic and Health Survey 1998* (pp. 89–96)*.* Calverton, MD: National Council for Population and Development (NCPD), Central Bureau of Statistics (CBS) (Office of the Vice President and Ministry of Planning and National Development) [Kenya], and Macro International Inc. (MI).

Kinzer, S. (2007). After So Many Deaths, Too Many Births. *New York Times* 11.02.2007. URL: http://www.nytimes.com/2007/02/11/weekinreview/11kinzer.html?_r=0.

Kögel, T., & Prskawetz, A. (2001). Agricultural Productivity Growth and Escape from the Malthusian Trap. *Journal of Economic Growth, 6*, 337–357.

Komlos, J., & Artzrouni, M. (1990). Mathematical Investigations of the Escape from the Malthusian Trap. *Mathematical Population Studies, 2,* 269–287.

Korotayev, A., & Khaltourina D. (2006). *Introduction to Social Macrodynamics: Secular Cycles and Millennial Trends in Africa*. Moscow: KomKniga/URSS.

Korotayev, A., Malkov A., & Khaltourina D. (2006*a*). *Introduction to Social Macrodynamics: Compact Macromodels of the World System Growth*. Moscow: URSS.

Korotayev, A., Malkov A., & Khaltourina D. (2006*b*). *Introduction to Social Macrodynamics: Secular Cycles and Millennial Trends*. Moscow: KomKniga/URSS.

Korotayev, A., & Zinkina, J. (2011). Egyptian Revolution: A Demographic Structural Analysis. *Entelequia. Revista Interdisciplinar, 13*, 139–169.

Korotayev, A., Zinkina, J., Kobzeva, S., Bogevolnov, J., Khaltourina, D., Malkov, A., & Malkov S. (2011). A Trap at the Escape from the Trap? Demographic-Structural Factors of Political Instability in Modern Africa and West Asia. *Cliodynamics: The Journal of Theoretical and Mathematical History, 2*(2), 276–303.

Lucas, R. (1998). On the Mechanics of Economic Development. *Journal of Monetary Economics, 22*, 3–42.

Maddison, A. (2010). *World Population, GDP and Per Capita GDP, A.D. 1–2008.* URL: www.ggdc.net/maddison.

Malthus, T. (1978 [1798]). *Population: The First Essay.* Ann Arbor, MI: University of Michigan Press.

Martorell, R., Kettel Khan, L., Hughes, M. L., & Grummer-Strawn, L. M. (2000). Obesity in women from developing countries. *European Journal of Clinical Nutrition, 54*, 247–252.

Mitchell, D., & Bareku M. (2012). The Tanzania Tobacco Sector: How Market Reforms Succeeded. In A. A. Ataman (Ed.). *African Agricultural Reforms. The Role of Consensus and Institutions* (pp. 271–290). Washington, DC: World Bank.

Moreland, S. (2006). *Egypt's Population Program: Assessing 25 Years of Family Planning*. Washington, DC: USAID.

Muindi, M., & Bicego, G. (1999). Aids and Other Sexually Transmitted Diseases. In *Kenya Demographic and Health Survey 1998* (pp. 127–160). Calverton, MD: National Council for Population and Development (NCPD), Central Bureau of Statistics (CBS) (Office of the Vice President and Ministry of Planning and National Development) [Kenya], and Macro International Inc.

Naiken, L. (2002). *FAO Methodology for Estimating the Prevalence of Undernourishment*. Paper Presented at International Scientific Symposium on Measurement and Assessment of Food Deprivation and Undernutrition, Rome, Italy. URL: www.fao.org.

National Bureau of Statistics [Tanzania] & Macro International Inc. (2000). *Tanzania Reproductive and Child Health Survey 1999*. Calverton, MD: National Bureau of Statistics and Macro International Inc.

National Bureau of Statistics (NBS) [Tanzania] & ORC Macro. (2005). *Tanzania Demographic and Health Survey 2004–05*. Dar es Salaam: National Bureau of Statistics and ORC Macro.

National Bureau of Statistics (NBS) [Tanzania] & ICF Macro. (2011). *Tanzania Demographic and Health Survey 2010*. Dar es Salaam: NBS and ICF Macro.

National Council for Population and Development (NCPD), Central Bureau of Statistics (CBS) (Office of the Vice President and Ministry of Planning and National Development [Kenya]), & Macro International. (1994). *Kenya Demographic and Health Survey 1993*. Calverton, MD: NCPD, CBS, and MI.

National Council for Population and Development, Institute for Resource Development, & Macro International. (1994). *Kenya Demographic and Health Survey 1993*. Calverton, MD: NCPD, CBS, and MI.

Nefedov, S. A. (2004). A Model of Demographic Cycles in Traditional Societies: The Case of Ancient China. *Social Evolution & History*, *3*(1), 69–80.

Nelson, R. R. (1956). A theory of the low level equilibrium trap in underdeveloped economies. *American Economic Review*, *46*, 894–908.

Ngallaba, S., Kapiga, S. H., Ruyobya, I., & Boerma, J. T. (1993). *Tanzania Demographic and Health Survey 1991/1992*. Dar es Salaam: Bureau of Statistics.

Opiyo, C., Omolo C., & Imbwaga A. (2010). Infant and Child Mortality. In *Kenya Demographic and Health Survey 2008–2009* (pp. 103–111). Calverton, MD: Kenya National Bureau of Statistics and ICF Macro.

Otieno, F., & Omolo C. (2004). Infant and Child Mortality. In *Kenya Demographic and Health Survey 2003* (pp. 114–122). Calverton, MD: Central Bureau of Statistics (CBS) [Kenya], Ministry of Health (MOH) [Kenya], and ORC Macro.

Pereira, A. S. (2006). *When Did Modern Economic Growth Really Start? The Empirics of Malthus to Solow.* Vancouver, BC: University of British Columbia.

Phillips, J. F., Stinson, W.S., Bhatia, S., Rahman, M., & Chakraborty, J. (1982). The demographic impact of the family planning – health services project in Matlab, Bangladesh. *Studies in Family Planning 13*(5), 131–140.

Roth, E. A., Franklin, E. (2005). The Social, Health, and Economic Consequences of Pastoral Sedentarization in Marsabit District, Northern Kenya. In E. Fratkin, E. A. Roth (Eds.). *As Pastoralists Settle. Social, Health, and Economic Consequences of Pastoral Sedentarization in Marsabit District, Kenya* (pp. 1–28). New York, NY: Kluwer.

Statistics Department [Uganda] & Macro International Inc. (1996). *Uganda Demographic and Health Survey, 1995.* Calverton, Maryland: Statistics Department [Uganda] and Macro International Inc.

Steinmann, G., & Komlos, J. (1988). Population Growth and Economic Development in the Very Long Run: A Simulation Model of Three Revolutions. *Mathematical Social Sciences, 16*, 49–63.

Steinmann, G., Prskawetz, A., & Feichtinger, G. (1998). A Model on the Escape from the Malthusian Trap. *Journal of Population Economics, 11*, 535–550.

Turchin, P. (2003). *Historical Dynamics: Why States Rise and Fall*. Princeton, NJ: Princeton University Press.

Turchin, P. (2005*a*). Dynamical Feedbacks between Population Growth and Sociopolitical Instability in Agrarian States. *Structure and Dynamics, 1*.

Turchin, P. (2005*b*). *War and Peace and War: Life Cycles of Imperial Nations.* New York, NY: Pi Press.

Turchin, P., & Korotayev, A. (2006). Population Density and Warfare: A Reconsideration. *Social Evolution & History, 5*(2), 121–158.

Turchin, P., & Nefedov, S. (2009). *Secular Cycles.* Princeton, NJ: Princeton University Press.

Uganda Bureau of Statistics (UBOS) & ORC Macro. (2001). *Uganda Demographic and Health Survey 2000–2001.* Calverton, MD: UBOS and ORC Macro.

Uganda Bureau of Statistics (UBOS) & Macro International Inc. (2007). *Uganda Demographic and Health Survey 2006.* Calverton, MD: UBOS and Macro International Inc.

Uganda Bureau of Statistics (UBOS) & ICF International Inc. (2012). *Uganda Demographic and Health Survey 2011.* Kampala, Uganda: UBOS and Calverton, Maryland: ICF International Inc.

Usher, D. (1989). The Dynastic Cycle and the Stationary State. *The American Economic Review, 79*, 1031–1044.

UN Population Division. (2012). *World Urbanization Prospects: The 2011 Revision, CD-ROM Edition.* New York, NY: United Nations, Department of Economic and Social Affairs, Population Division.

UN Population Division. (2014). *World Population Prospects.* URL: http://esa.un.org/unpd/wpp/.

WHO. (2008). Sharing the burden of sickness: mutual health insurance in Rwanda. *Bulletin of World Health Organization, 86*(11), 823–824.

Wood, J. W. (1998). A Theory of Preindustrial Population Dynamics: Demography, Economy, and Well-Being in Malthusian Systems. *Current Anthropology, 39*, 99–135.

World Bank. (2014). *World Development Indicators Online.* Washington, DC: World Bank. URL: http://data.worldbank.org/indicator.

Zinkina, J., & Korotayev, A. (2013). Urbanization Dynamics in Egypt: Factors, Trends, Perspectives. *Arab Studies Quarterly, 35*(1), 20–38.

**Appendix 1. Tanzanian case**

Tanzania constitutes a rather special case, as the stagnation of per capita food consumption was observed in this country against the background of not only the GDP per capita growth, but also some growth of labor productivity in agriculture (see Fig. 5 above). This seems to be accounted for by the point that in Tanzania recent years evidenced a rapid growth of production of such agricultural products as tea (see Fig. 1.1) or tobacco[13] (see Fig. 1.2) where production take place to a considerable extent at larger agricultural enterprises, whereas the production of such staple foods as cassava (see Fig. 1.3), where the majority of agricultural population is concentrated, stagnated or even declined.

**Fig. 1.1.** Dynamics of tea production in Tanzania (tons)

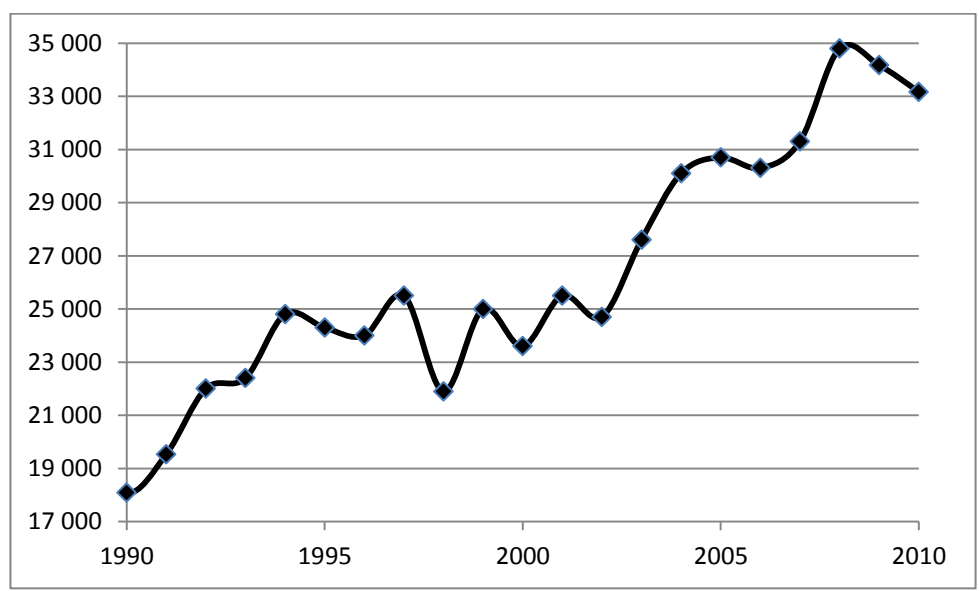

*Data source*: FAO, 2014.

---

[13] On the recent success story in the Tanzanian tobacco cultivation see in particular Mitchell & Baregu, 2012.

**Fig. 1.2.** Dynamics of unmanufactured tobacco production in Tanzania (tons)

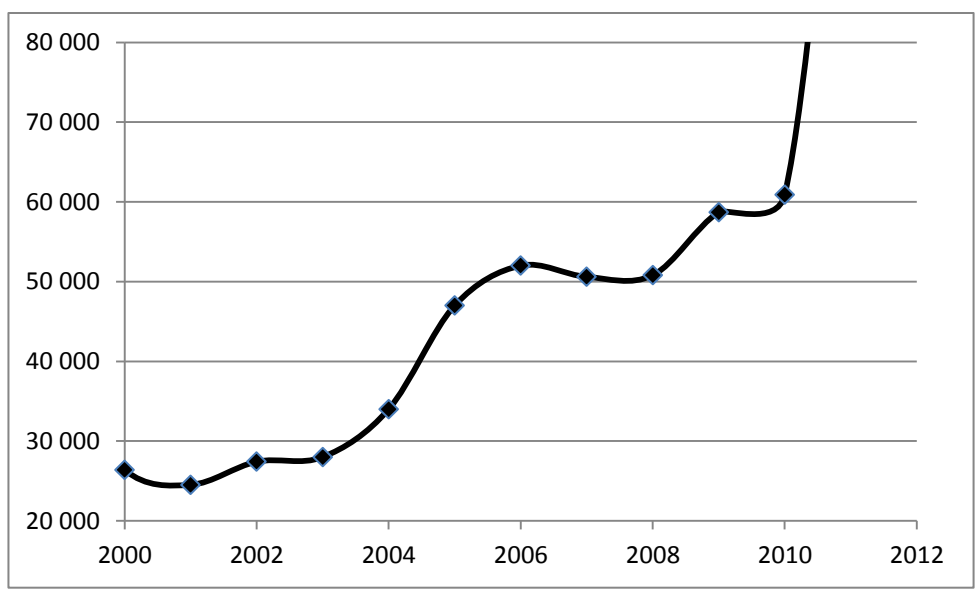

*Data source*: FAO, 2014.

**Fig. 1.3.** Dynamics of cassava production in Tanzania (tons)

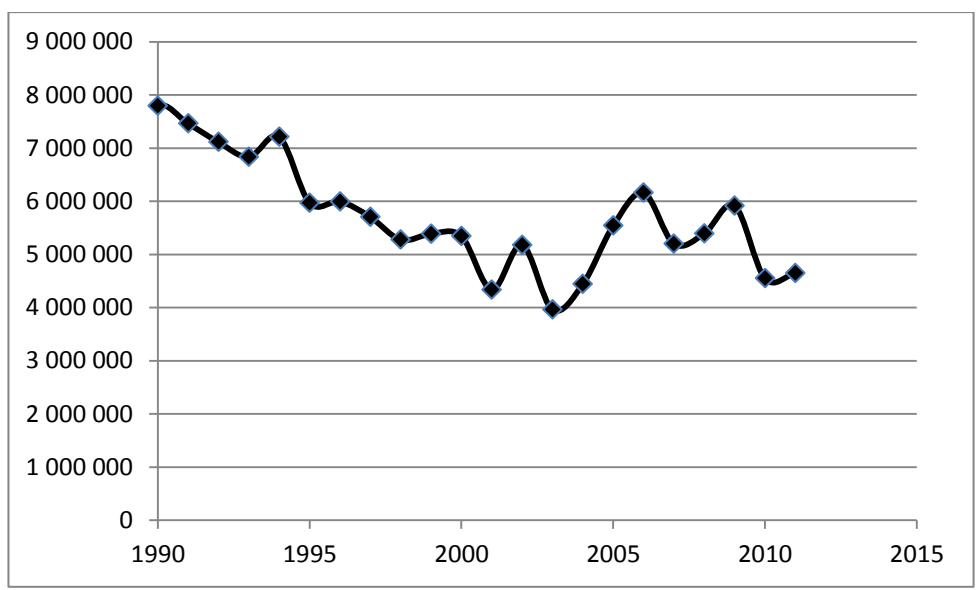

*Data source*: FAO, 2014.

Born in Moscow, **Andrey Korotayev** attended Moscow State University, where he received a B.A. degree in 1984 and an M.A. in 1989. He earned a Ph.D. in 1993 from Manchester University, and in 1998 a Doctor of Sciences degree from the Russian Academy of Sciences. He is currently the Head of the Laboratory of Monitoring of the Risks of Sociopolitical Destabilization of the National Research University Higher School of Economics and a Senior Research Professor at the Center for Big History and System Forecasting of the Institute of Oriental Studies as well as in the Institute for African Studies of the Russian Academy of Sciences. In addition, he is a Senior Research Professor of the Laboratory of Political Demography and Social Macrodynamics of the Russian Academy of National Economy and Civil Administration, as well as a Full Professor of the Faculty of Global Studies of theMoscow State University. He is co-editor of the journals *Social Evolution & History* and *Journal of Globalization Studies*, as well as *History & Mathematics* and *Evolution* almanac. Together with Askar Akayev and George Malinetsky he is a coordinator of the Russian Academy of Sciences Program "System Analysis and Mathematical Modeling of World Dynamics". Korotayev is a laureate of the Russian Science Support Foundation in "The Best Economists of the Russian Academy of Sciences" nomination (2006). In 2012 he was awarded with the Gold Kondratieff Medal by the International N. D. Kondratieff Foundation. His hard mail is 20 Myasnitskaya, Moscow 101000, Russia. E-mail: akorotayev@gmail.com. Phone: +7 917 517 8034

**Julia Zinkina** has her PhD in History and works as a Research Fellow at the Laboratory for Monitoring Sociopolitical Destabilization Risks, Higher School of Economics, and Institute for African Studies (Russian Academy of Sciences), Moscow, Russia. In addition, she is a Research Professor of the Laboratory of Political Demography and Social Macrodynamics of the Russian Academy of National Economy and Civil Administration. Her hard mail is 30/1 Spiridonovka, Moscow 123001, Russia. E-mail: juliazin@list.ru. Phone: +7 903 964 3809